\documentclass[8.5pt,twoside,twocolumn]{article}

\usepackage{setspace}

\usepackage{scrextend}

\usepackage{anysize}
\marginsize{1.25cm}{1.25cm}{1.2cm}{1.15cm}

\usepackage{tikz}
\usepackage{pgfplots}

\usepackage[format=plain,font=small]{caption} 

\usepackage{epstopdf, epsfig}
\usepackage{float}
\usepackage{booktabs}
\usepackage{amssymb}
\usepackage{pgfplotstable}
\usepackage{amsmath}
\usepackage{graphicx}
\usepackage{pgfplots}
\usepackage{tikz}
\usetikzlibrary{snakes}
\usetikzlibrary{patterns}
\usepackage{xcolor}
\usepgfplotslibrary{colormaps}
\usepackage{lineno}
\usepackage{hyperref}

\newcommand{\La}{{\rm{La}}}

\newcommand{\Ca}{{\rm{Ca}}}
\newcommand{\vzero}{\mathbf{0}}

\newcommand{\vn}{\mathbf{n}}

\newcommand{\ve}{\mathbf{e}}
\newcommand{\vv}{\mathbf{v}}
\newcommand{\vx}{\mathbf{x}}

\newcommand{\id}{\mathcal I}

\newcommand{\disp}{L}
\newcommand{\bdisp}{\bar{L}}
\newcommand{\V}{\mathcal{V}}

\newcommand{\SigmaN}{\Sigma_{n}}

\newcommand{\Rmen}{R_{m}}
\newcommand{\bRmen}{\bar{R}_{m}}

\newcommand{\bR}{\bar{R}}
\newcommand{\vnabla}{\mathbf{\nabla}}

\newcommand{\figref}[1]{Fig.~\ref{#1}}

\newcommand{\tabref}[1]{Table~\ref{#1}}

\newcommand{\stress}{\mathbf{\tau}}

\newcommand{\dd}{{\rm{d}}}

\newcommand{\rojo}[1]{}

\pgfplotsset{compat=newest}
\usepgflibrary{decorations.markings}
\usepgflibrary{decorations.shapes}
\definecolor{green}{rgb}{0.0,.6,0.0}
\definecolor{migris}{rgb}{.85,.85,.85}%

\tikzset{dashed/.style={dash pattern=on 3pt off 1pt}}
\tikzset{dashdot/.style={dash pattern=on .4pt off 3pt on 4pt off 3pt}}
\tikzset{,
	MyPersp1/.style={scale=1.8,x={(-0.8cm,-0.4cm)},y={(0.8cm,-0.4cm)},z={(0cm,1cm)}},
	MyPoints/.style={fill=white,draw=black,thick}
		}
\pgfplotsset{
    major tick length=.2cm,
    minor tick length=.1cm,
        max space between ticks=10,
	axis on top,
	xtick align=center,
	ytick align=center,
	xlabel near ticks,
	ylabel near ticks,
         legend cell align=left,
	width={0.9\textwidth},
    	height={0.0\textwidth},
	scale only axis,
	legend style={font=\tiny},
	title style={font=\footnotesize},
	every axis/.append style={font=\footnotesize},
	ticklabel style={font=\footnotesize},
	xlabel style={font=\footnotesize},
	ylabel style={font=\footnotesize},
	xticklabel style={font=\footnotesize},
	every axis plot/.append style={line width=.75pt,mark size=3pt},
	legend style={row sep=-4pt,legend cell align=left, align=left, legend plot pos=left,draw=none,fill=none},
	extra tick style={
        		tick align=outside,
        		tick pos=left,
        		grid style={dotted,black},
    	},
	}

\begin{document}

\thispagestyle{plain}

\twocolumn[
  \begin{@twocolumnfalse}
\noindent\LARGE{\textbf{Raydrop : a universal droplet generator based on a non-embedded co-flow-focusing}}
\vspace{0.6cm}

\noindent\large{\textbf{Adrien Dewandre$^{\ast \dag}$, Javier Rivero-Rodriguez$^{\dag}$, Youen Vitry, Benjamin Sobac and
Benoit Scheid}}\vspace{0.5cm}

\begin{addmargin}{1cm}
\noindent \small{Most commercial microfluidic droplet generators rely on the planar flow-focusing configuration implemented in polymer or glass chips. The planar geometry, however, suffers from many limitations and drawbacks, such as the need of specific coatings or the use of dedicated surfactants, depending on the fluids in play. On the contrary, and thanks to their axisymmetric geometry, glass capillary-based droplet generators are a priori not fluid-dependent. Nevertheless, they have never reached the market because their assembly requires art-dependent and not scalable fabrication techniques. Here we present a new device, called Raydrop, based on the alignment of two capillaries immersed in a pressurized chamber containing the continuous phase. The dispersed phase exits one of the capillaries through a 3D-printed nozzle, placed in front of the extraction capillary for collecting the droplets. This non-embedded implementation of an axisymmetric flow-focusing is referred to {\it co-flow-focusing}. Experimental results demonstrate the universality of the device in terms of the variety of fluids that can be emulsified, as well as the range of droplet radii that can be obtained, without neither the need of surfactant nor coating.  Additionally, numerical computations of the Navier-Stokes equations based on the quasi-steadiness assumption are shown to correctly predict the droplet radius in the dripping regime and the dripping-jetting transition when varying the geometrical and fluid parameters. The monodispersity ensured by the dripping regime, the robustness of the fabrication technique, the optimization capabilities from the numerical modeling and the universality of the configuration confer to the Raydrop technology a very high potential in the race towards high-throughput droplet generation processes.}
\end{addmargin}

\vspace{0.5cm}
 \end{@twocolumnfalse}
 ]

\section{Introduction}
\label{Introduction}

\footnotetext{\textit{
$\ast$ TIPs Lab, Universit\'e libre de Bruxelles, Brussels, Belgium. E-mail: adrien.dewandre@ulb.ac.be\\
$\dag$ The two first authors contributed equally to this work}
}

\begin{figure}[ht!] 
        \centering 
        \includegraphics[width=1\columnwidth]{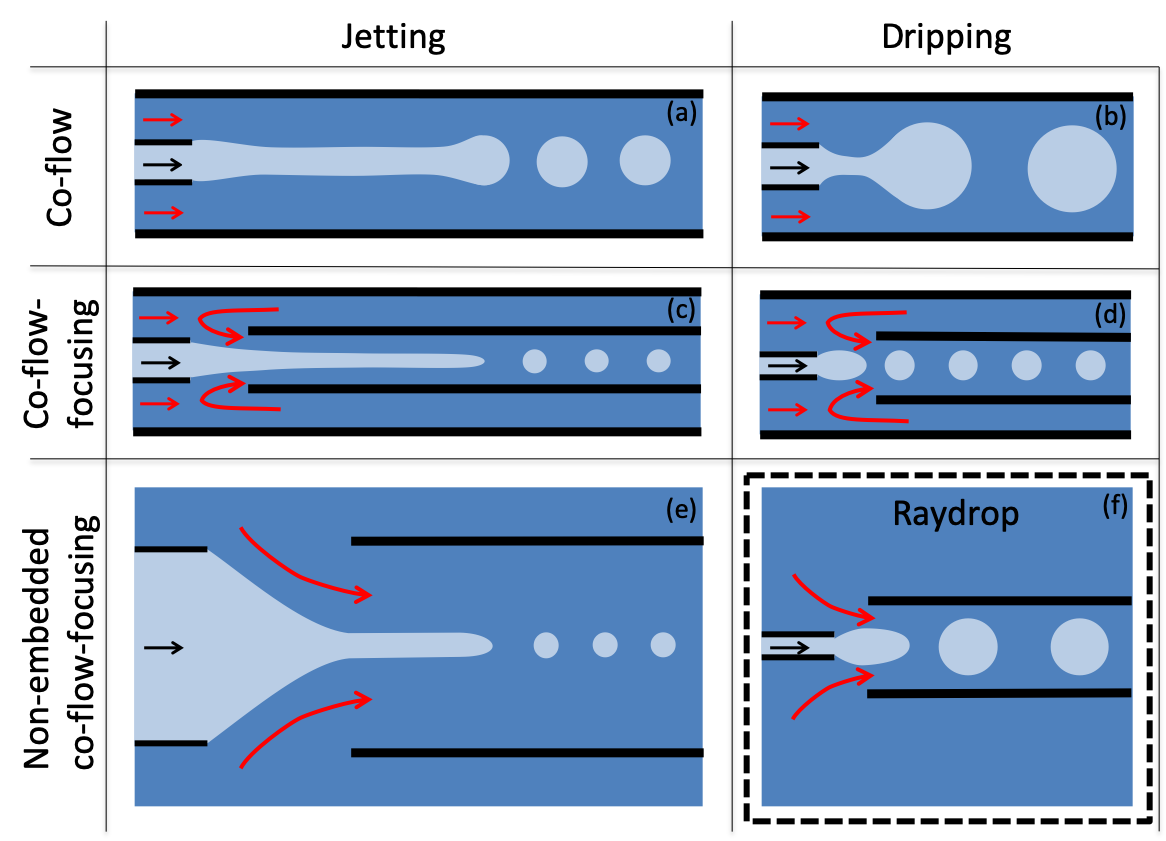}
        \caption{Capillary-based axisymetric designs of droplet generators. (a) and (b) jetting and dripping of a co-flow, respectively~\cite{Guillot07};  (c) and (d) jetting and dripping of a {\it co-flow-focusing} in an embedded geometry, respectively~\cite{Utada05};  (e) {\it co-flow-focusing} in a non-embedded geometry, enforcing the jetting regime~\cite{Evangelio16}, also referred to as tip-streaming; (f) Raydrop : {\it co-flow-focusing} in a non-embedded geometry, enforcing the dripping regime.}
        \label{fig:regimes}
\end{figure}

In recent years, droplet microfluidics~\cite{Gunther06,Teh08,Tran13,Anna16,Shang17} has become an important tool for many different applications, including fundamental studies on emulsification~\cite{Bremond12}, crystalization~\cite{Candoni19} and  chemical reaction~\cite{Song06}, temperature-controlled tensiometry~\cite{Lee17}, molecules encapsulation~\cite{Conchouso14}, particle synthesis~\cite{Nunes13}, biochemical assays~\cite{Baccouche17}, immunoassays~\cite{Ali-Cherif12}, digital PCR~\cite{Zhang16}, directed evolution~\cite{Agresti10}, drug discovery~\cite{Damiati18}, single-cell analysis~\cite{Klein15} or cell and gene manipulations~\cite{Brouzes09,Ryckelynck15}. Nowadays, all the commercially available and most of lab-made droplet generators are based on a flow-focusing technology implemented in rectangular microchannels fabricated by lithography, and made of polydimethylsiloxane (PDMS), polymers or glass. However, this planar configuration has many limitations, mostly due to the contact between the walls of the microchannels and both phases at the junction, making laborious and often ephemeral wettability treatments of these walls necessary. On the contrary, due to their axisymmetric configuration, glass capillary systems do not have this drawback since the dispersed phase is never in contact with the walls of the outer capillary~\cite{Guillot07}. Yet their widespread use is limited by the difficulty to implement this technology in an easy-to-use device. And even if co-flow configurations have been set up using commercially available components~\cite{Serra07}(see Fig.~\ref{fig:regimes}(a),(b)), these devices produce large droplets ($>100~\mu$m) at low flow rates and are unable to generate small ($< 100~\mu$m) and monodisperse droplets at high flow rates ($>1$~kHz), as realised in planar flow focusing configuration. Furthermore, the centring of the capillary into an outer flow capillary is challenging. Even though the flow focusing configuration has also been designed using glass capillaries by inserting two circular capillaries into a square outer flow capillary, which greatly simplifies centring of the capillaries~\cite{Utada05}(see Fig.~\ref{fig:regimes}(c),(d)), it remains three main limitations. Firstly, the restrictions necessary to obtain a focusing effect at the tip of the inner injection and extraction capillaries are obtained by pulling and breaking a heated glass capillary, a very art-dependent technique limiting the use of this design in large-scale microfluidic applications.  Secondly, the confined space for flowing the continuous phase around the inner capillaries limits its flow rate, hence the throughput, in the droplet generation process, particularly with highly viscous fluids. And finally, the interfacing with the tubing carrying the fluids is typically realised by gluing hypodermic needles over the embedded capillaries previously glued on a glass slide. This fabrication process not only limits the reuse of the device because it can not be disassembled and properly cleaned, but also make it not suitable for large-scale production. Alternative procedures have been developed to improve the usability of embedded capillary devices~\cite{Benson13,Bandulasena19}, but none can overcome all these limitations. 

A configuration proposed by Evangelio {\it et al.}~\cite{Evangelio16} has revealed a promising alternative by placing the extraction tube without any surrounding confinement in front of the injection tube (see Fig.~\ref{fig:regimes}(e)). The continuous phase is therefore accelerated to the extraction tube, thus creating a pressure drop, similar to a Venturi tube, even though viscous forces are dominant in this case. The authors have exploited this pressure drop to create a stretched jet of the dispersed phase, that further destabilizes into micro-bubbles or droplets, a mechanism which is referred to as the ``tip streaming". However, their system exclusively works in the jetting regime, which does not intrinsically guarantee the droplet monodispersity associated to the dripping regime, as realised in embedded glass capillaries~\cite{Utada05}.
In this work, we consider a new configuration based on the previous system but in which the dripping regime is enforced.

Our device relies on the alignment of two glass capillaries inside a pressurised chamber, similarly to Evangelio {\it et al.}~\cite{Evangelio16}.  A 3D-printed micro-nozzle is additionally connected at the tip of the injection capillary (see Fig.~\ref{fig:dropbox_full}(d)), enforcing the dripping of small droplets as in Utada {\it et al.}~\cite{Utada05}. This non-embedded design presents both the characteristics of a co-flow (axisymetric geometry) and a flow-focusing (dramatic local accelerations of the continuous phase), and is thereby called non-embedded {\it co-flow-focusing} (see Fig.~\ref{fig:regimes}(f)). The technological breakthrough of this design is two-folds. Firstly, it enables high-throughput generation of monodisperse droplets, intrinsic to the dripping regime, for a wide variety of fluids, thanks to the fact that the continuous phase is not confined before entering the extraction capillary, allowing for flushing very viscous continuous phases. Secondly, it takes benefit of specific fabrication techniques and materials compatible with a large-scale production of the device, while ensuring a very precise and reproducible alignment of the two capillaries in the chamber. Additionally, the device is made plug-and-play thanks to the standard connections and the possibility to easily assemble and disassemble all parts for cleaning. 

Besides the axisymmetric configuration proposed by Evangelio {\it et al.}~\cite{Evangelio16} in the jetting mode, one should mention the recent experiment performed by Cruz-Mazo {\it et al.}~\cite{Cruz-Mazo16} that exploits the axisymmetric flow-focusing in the dripping mode. To our knowledge, this is the closest configuration to the one of the Raydrop. Yet these authors used exclusively gas as continuous phase and found that the liquid drop size is independent of the flow rate of the dispersed phase, a feature that suggests a quasi-static behavior, as exploited in this work.

The ultimate aim in droplet generation being to predict the size of droplets, initial theoretical works have relied on linear stability analysis of the co-flow configuration and found that the dripping and jetting regimes exhibit the main features of absolutely and convectively unstable flows, respectively~\cite{Guillot07,Guillot08}. However, Cordero {\it et al.}~\cite{Cordero11} have shown that the frequency selection in the dripping regime is not ruled by the absolute frequency predicted by the stability analysis. 
Thus recognizing the nonlinear behaviour of droplet formation in the dripping regime, many numerical studies have been performed in the co-flow configuration using the Navier-Stokes equations together with diverse methods to describe the interfacial dynamics, as for instance the finite volume method combined with continuous-surface-force method~\cite{Richards95}, front tracking method~\cite{Hua07} or level-set method~\cite{Lan15}. For the flow-focusing configuration in a cross-junction microchannel, one can mention the diffuse interface or phase-field method~\cite{Bai17}, or the work by Wu {\it et al.}~\cite{Wu08} using three-dimensional lattice Boltzmann simulations. Nevertheless, the computing heaviness of three-dimensional simulations and the number of geometrical parameters involved in a planar flow-focusing configuration have prevented so far systematic and complete numerical analysis of the drop formation in such a configuration. Using finite element method (FEM) with adaptive meshing in a diffuse-interface framework, Zhou {\it et al.}~\cite{Zhou06} have simulated an axisymmetric flow-focusing configuration, yet involving a large number of geometrical parameters. The non-embedded {\it co-flow-focusing} presented here reduces the configuration to the minimum number of independent parameters, then offering the possibility of an exhaustive inspection. Yet the task is still huge and some additional simplifications are needed. 

Several authors have recognised the quasi-static behavior of the dripping mode. For instance, Garstecki {\it et al.}~\cite{Garstecki05} claimed that the quasi-static character of their data collapse forms the basis for controlled, high-throughput generation of monodisperse fluid dispersions. In the step-emulsification configuration, Li {\it et al.}~\cite{Li15} solved the quasi-static shape of an elongated drop using the 1D equations of the Hele-Shaw cell. Inspired by these quasi-static behaviours, and thinking about the {\it co-flow-focusing} configuration, which is intrinsically axisymmetric, the analogy with the pendant droplet becomes evident. In this context, the quasi-static assumption applies, provided the liquid is injected at a sufficiently low flow rate into the droplet in formation. Boucher {\it et al.}~\cite{Boucher75} for instance have shown that the dripping of a pendant droplet coincides with a folding bifurcation, indicating the maximum drop volume above which surface tension cannot sustain the drop weight anymore. But gravity can obviously be replaced by other forces, such as surface forces~\cite{Valet18}, or electric forces~\cite{Basaran90}. We claim in this paper that the quasi-static approach can be adopted analogously to the pendant drop method, provided the flow rate of the dispersed phase is small enough and that the role of gravity is essentially played by the hydrodynamic forces acting on both phases and originated by the imposed continuous flow rate. As mentioned above, this situation has been experimentally considered by Cruz-Mazo {\it et al.}~\cite{Cruz-Mazo16}, but only in the case of air as continuous phase. We here consider all situations in the quasi-static approach, from zero viscosity ratio corresponding to inviscid droplets, to infinite viscosity ratio corresponding to highly viscous droplets.

In the context of simulating multi-components flows, interface tracking methods, such as finite element methods (FEM) with moving mesh algorithm, are very accurate for simulating the onset of droplet break-up but have difficulties in simulating through and past the transitions as it requires cut-and-connect procedures for the mesh~\cite{Cristini04}. Assuming that the volume of the dispersed phase preceding the pinch-off corresponds exactly to the volume of the droplet after break-up, there is in fact no need to simulate the dynamics beyond the transition if one only wants to determine the volume of the generated droplet. Following this approach, Martinez-Calvo {\it et al.}~\cite{Martinez20} have applied FEM to investigate the break-up of liquid threads and determine the volume of satellite droplets, with an impeccable accuracy. They have solved the Navier-Stokes equations using the finite element software Comsol, and more precisely, the weak form PDE and the Arbitrary Lagrangian-Eulerian (ALE) modules. Note that the jet breaking is intrinsically transient as it results from the Rayleigh-Plateau instability but the droplet generation, as mentioned above, can be quasi-static in the limit of low dispersed flow rate. Van Brummelen {\it et al.}~\cite{VanBrummelen01} having shown that solving the transient Navier-Stokes equation to obtain the steady solution is often inefficient, Rivero-Rodriguez {\it et al.}~\cite{Rivero20} have developed the Boundary Arbitrary Lagrangian-Eulerian (BALE) method to facilitate the solution of steady Navier-Stokes equations with free surfaces. Rivero-Rodriguez and Scheid~\cite{Rivero18,Rivero19} have then applied this method to study the dynamics of deformable and off-centered bubbles in microchannels, allowing for exhaustive parametric analysis, taking the advantage of continuation methods suitable for tracking stationary solutions. The same approaches are undertaken in the present paper, offering an unprecedented parametric analysis in the context of flow-focusing.

After detailing the Raydrop configuration in section~\ref{Materials and methods}, we show in section~\ref{Experiments} the experimental measurements of droplet size and frequency for two different geometries, with the aim to identify experimentally the dripping to jetting transition when increasing both the dispersed and the continuous flow rates. Section~\ref{Modelling} is dedicated to the model description and validation in regards to the experimental results, and especially the relevance of the quasi-static approach for determining the droplet radius in the dripping regime in the limit of small dispersed flow rate. Section~\ref{Parametric analysis} focuses on the quasi-static approach, first for inviscid droplets, then for viscous ones, followed by a parametric analysis of the geometrical parameters, and a discussion on the role of inertia for properly predicting the dripping to jetting transition when increasing the continuous flow rate. Conclusions and perspectives are given in section~\ref{Conclusions}.

\section{Materials and methods}
\label{Materials and methods}
\begin{figure*}[ht!] 
        \centering 
        \includegraphics[width=\textwidth]{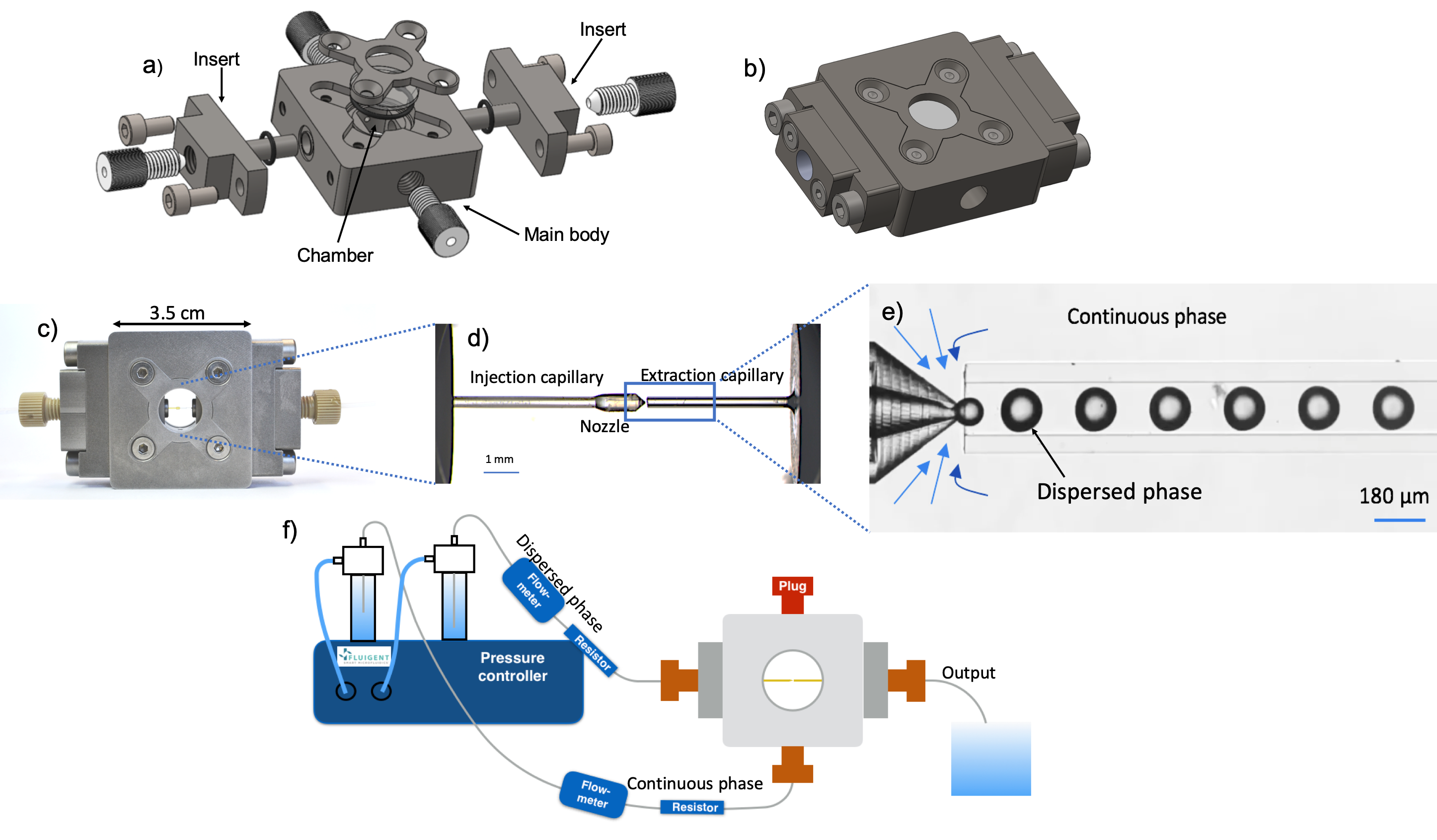}
        \caption{\fontsize{9}{9}\selectfont (a) Exploded view and (b) assembled view of the Raydrop. (c) The Raydrop with the injection and extraction glass capillaries. (d) Zoom through the top window of the two capillaries aligned in the chamber filled with the continuous phase. (e) Zoom on the droplet generation area. The 3D-printed nozzle connected to the injection capillary carries the dispersed phase while the extraction capillary collects droplets of the dispersed phase entrained by the continuous phase. At the entrance of the extraction capillary, the continuous phase is dramatically accelerated because of the change of section and squeezes the dispersed phase, resulting in the droplets formation. (f) Setup used for the production of droplets using the Raydrop. The flows are controlled using a pressure controller and the flow rates are measured using flow-meters. The optical materials allowing the observation of droplets by transmission through the windows, i.e. light source, microscope and high-speed camera, are not shown.}
 	\label{fig:dropbox_full} 
\end{figure*}

The droplet generator Raydrop is made of a metallic chamber filled with the continuous phase, in which two inserts supporting the glass capillaries are introduced on the lateral sides in such a way that the capillaries are perfectly aligned and almost in contact at the centre of the chamber, as shown in Fig.~\ref{fig:dropbox_full}. 

Two glass windows on the top and bottom faces of the device close the chamber and allow the observation of the droplets. Leakage from the chamber is prevented by the use of O-rings seals on the windows and inserts.
A nozzle is printed in the photoresin IP-L  with a sub-micrometric resolution using a 3D-printer Photonic Professional GT from Nanoscribe company and then glued onto the tip of the injection glass capillary. Capillaries (Postnova Analytics) are coated with a polyimide film on the external diameter, ensuring a very high mechanical resistance. They are held into the inserts in such a way that for all combinations, alignment is guaranteed and a fixed gap between the nozzle and the extraction capillary is maintained. As demonstrated later, one can then change in a few minutes the couple of nozzle and extraction capillary in order to have access to other ranges of droplet radii, which represents a great advantage as compared to most of other glass capillary devices. Connections to the fluids inlets and outlets are ensured by standard Upchurch fittings directly screwed on the inserts for the connection to the capillaries and on the chamber for the continuous phase supply and bleed.

As shown in Fig.~\ref{fig:dropbox_full}(d), the input capillary supporting the nozzle provides the dispersed phase while the output capillary collects droplets carried by the continuous phase. At the entrance of the output capillary, the continuous phase pressurised in the chamber encounters a dramatic acceleration due to the change of section and therefore squeezes the dispersed phase flowing out of the nozzle, resulting in the formation of droplets (see Fig.~\ref{fig:dropbox_full}(e)). 

Fluids are injected in the device using a pressure controller (MFCS-EZ, Fluigent) and the flow rates are measured using flow-meters (Flow Unit, Fluigent), as shown in Fig.~\ref{fig:dropbox_full}(f). A high-speed camera (MotionPro Y3, IDT) operating up to 10000 frames per second is connected to a microscope with a 10x zoom to visualise the droplets. Recorded images are then processed with a Python script to detect the contour of the droplets and determine their size.

\section{Experiments}
\label{Experiments}
To demonstrate the operation of the device, we used two nozzles of tip radii $R_n = 15$ or $45~\mu$m, coupled with an extraction capillary of internal radius $R_e= 75$ or $225~\mu$m, respectively. The distance $H$ between the nozzle tip and the extraction capillary is 50~$\mu$m, the thickness $e$ of the walls of the extraction capillary is 50 or 75~$\mu$m and the external angle of the nozzle $\alpha$ taken perpendicularly from the main flow axis is 50$^\circ$. We used ultrapure water, obtained from a Sartorius water filtration system, as dispersed phase and light mineral oil, purchased from Sigma-Aldrich, as continuous phase. Both phases are filtered with PTFE 0.45~$\mu$m syringe filters. The interfacial tension of the bare interface between the two fluids was measured to be 49.5~mN/m using the pendant drop method (tensiometer Kruss DSA-100). Unlike glass chips relying on a flow-focusing junction, capillary-based devices do not necessitate the use of a surfactant to generate droplets, in both water in oil (W/O) and oil in water (O/W) cases. Even though surfactants can be used to stabilise the emulsion after the droplets are collected and stored, the literature mentions that its presence does not affect the droplet formation (see for instance Erb {\it et al.}~\cite{Erb11}). Indeed, adsorption times at the interface of the liquids are on a typical scale of tens of milliseconds whereas the droplet formation occurs in a few milliseconds. To validate this assumption in the case of our device, droplets generated with and without surfactant (Span 80, 2\% w/w in the continuous phase) have been compared, with no difference. Fluid properties and geometrical parameters for two different couples of nozzle and extraction capillary mentioned above are listed in \tabref{tbl:data}. Subscripts $d$ and $c$ refer to the dispersed and continuous phases, respectively, $\mu$ being the dynamic viscosity, $\gamma$ the interfacial tension and $\rho$  the density of the fluids.
\begin{table}[h]
\centering
\begin{tabular}{|lc|l|c|c|}
\hline
\multicolumn{2}{|c|}{Fluid properties}           & \multicolumn{3}{c|}{Geometrical parameters}   \\ \hline
\multicolumn{1}{|c}{}                     &      &                     & Couple 1      & Couple 2      \\ \hline
\multicolumn{1}{|l|}{$\mu_d$ (mPa.s)}     & 1    & $R_n$ ($\mu$m)        & 15         & 45         \\ \hline
\multicolumn{1}{|l|}{$\mu_c$ (mPa.s)}     & 23   & $R_e$ ($\mu$m)        & 75         & 225        \\ \hline
\multicolumn{1}{|l|}{$\gamma$ (mN/m)}     & 49.5 & $H$ ($\mu$m)         & 50 & 75 \\ \hline
\multicolumn{1}{|l|}{$\rho_d$ (kg/m$^3$)} & 1000 & $e$ ($\mu$m)         & \multicolumn{2}{c|}{75} \\ \hline
\multicolumn{1}{|l|}{$\rho_c$ (kg/m$^3$)} & 860  & $\alpha$ (deg)      & \multicolumn{2}{c|}{50} \\ \hline
\end{tabular}
 \caption{Fluid properties and geometrical parameters used in this work with water and oil for the dispersed and the continuous phases, respectively. ``Couple" refers to a specific combination of a nozzle and an extraction capillary.}
\label{tbl:data}
\end{table}

\begin{figure*}[ht!] 
        \centering 
        \includegraphics[width=\textwidth]{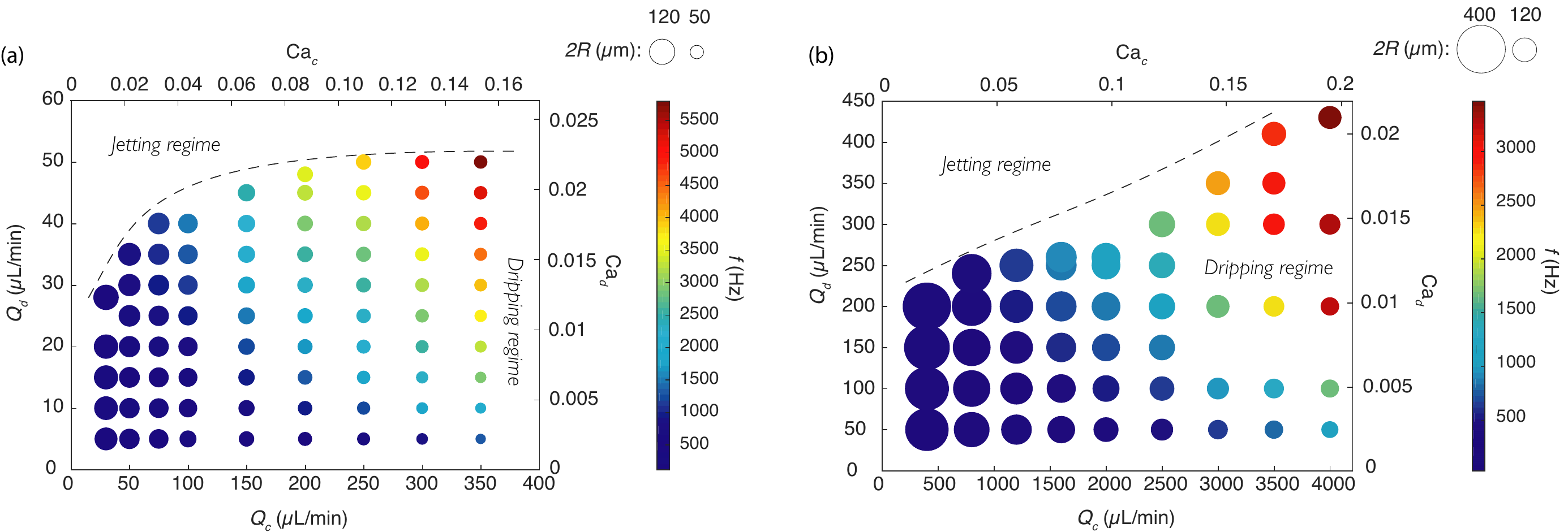}
        \caption{Size and production frequency of water droplets in mineral oil as a function of the continuous and dispersed flow rates $Q_c$ and $Q_d$, respectively, for the parameters given in \tabref{tbl:data}, with (a) couple 1 and (b) couple 2 for the geometries. The values of the capillary numbers for the dispersed and continuous phases, as defined in eqn~(\ref{Ca}) and with $\Ca_d=\bar Q \Ca_c$, are also given on the right and top axes, respectively.}
        \label{Fig:WO30mu} 
\end{figure*}

Considering the first couple of nozzle and extraction capillary (see dimensions in \tabref{tbl:data}), for each value of the oil flow rate $Q_c$ between 20 and 350~$\mu$L/min, the flow rate of the dispersed water $Q_d$ is increased until it reaches the dripping-jetting transition. The results are shown in Fig.~\ref{Fig:WO30mu}(a) where the size of the bullet is proportional to the size of the droplet and the colour scale indicates the frequency of droplets generation. 
We observe with the geometry of couple 1 that droplets are generated in a range of radius $R$ from 25 to 60~$\mu$m with a high monodispersity (CV$<$2\%). For a given $Q_d$, the droplet radius decreases by increasing $Q_c$, while for a given $Q_c$, it remains very stable as $Q_d$ is increased, with a maximum radius variation of 15\% at $ Q_c = 350 ~\mu$L/min. Additionally, as the system operates for $Q_d \gtrsim 200$\,$\mu$L/min, the dripping-jetting transition (dashed line in Fig.~\ref{Fig:WO30mu}(a)) reaches a plateau indicating that the dripping-jetting transition is only determined by the geometry in this region. Yet the droplet generation frequency increases to reach a maximum of 5500\,Hz in this case. We have also tested the sensitivity of the device to small geometrical modifications such as a small misalignment of both capillaries, namely of about a fraction of the nozzle tip radius, i.e., $<R_n/4$.  The influence of the distance $H$  between the nozzle tip and the extraction capillary has also been tested in the range 0 to 2.5 $R_n$. In both tests, droplet radii have been found similar to these obtained with the nominal values of the parameters. 

Now, for the geometry of couple 2 (see dimensions in \tabref{tbl:data}), droplets can be generated in a range of radius from 60 to 200~$\mu$m (see Fig.~\ref{Fig:WO30mu}(b)). As for couple 1, a dripping-jetting transition is observed when $Q_d$ is increased above a threshold value (dashed line), although a plateau value was not yet reached here for increasing $Q_c$ up to the maximum value allowed by the pressure controller. 

\begin{figure*}[ht!] 
        \centering 
        \includegraphics[width=0.85\textwidth]{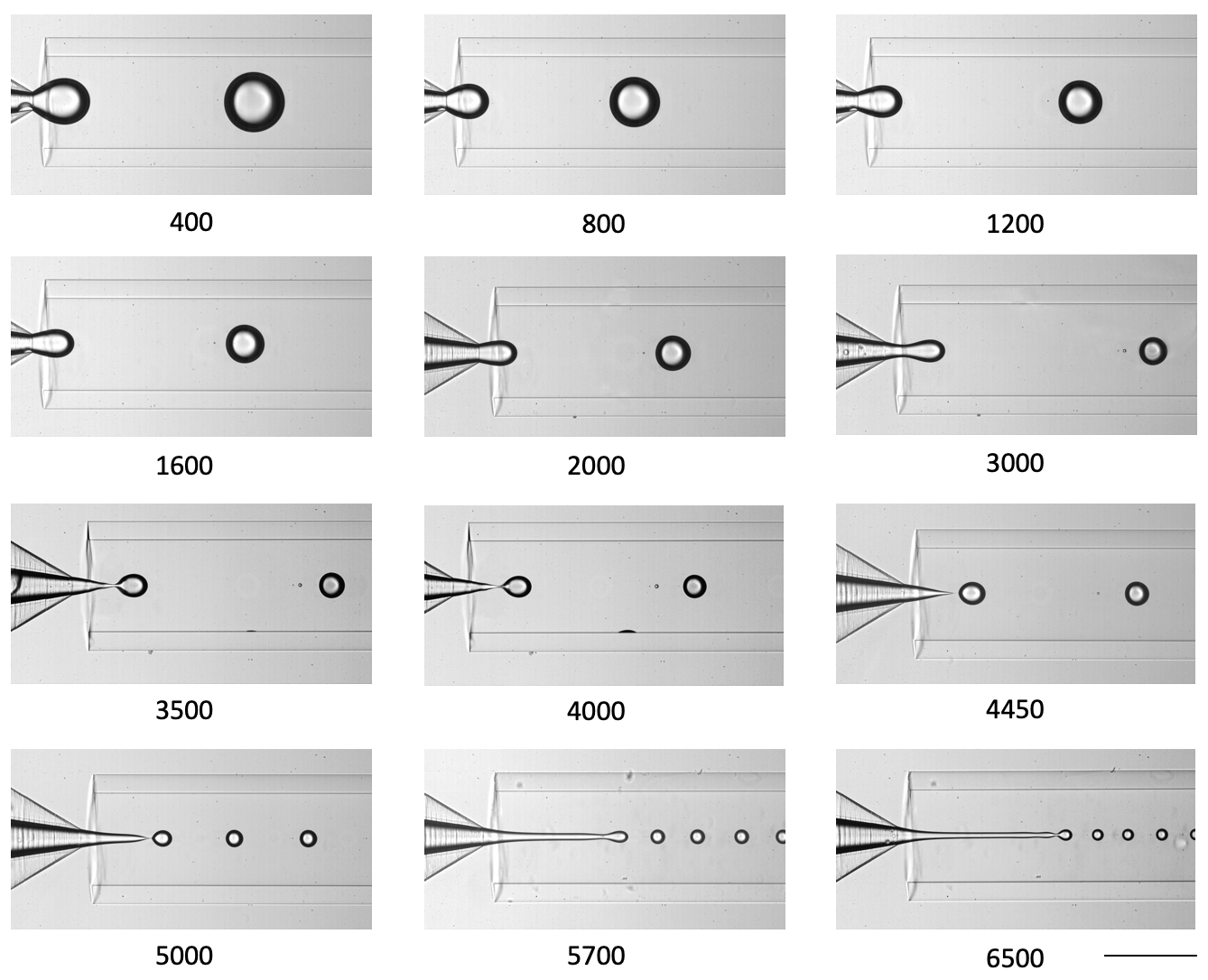}
        \caption{Water droplets in mineral oil using couple 2. $Q_d$ is fixed at 50~$\mu$L/min while the values of $Q_c$ are reported below each image. The scale bar is 450 $\mu$m. Capillary numbers as defined in eqn~(\ref{Ca}) and with $\Ca_d=\bar Q \Ca_c$, are $\Ca_d=0.0024$ and $\Ca_c=4.9\times10^{-5}Q_c$, with $Q_c$ in $\mu$L/min.}
        \label{Fig:DripJet} 
\end{figure*}

While only the dripping regime could be observed for couple 1, because of the limitation of the maximum pressure drop attainable, higher flow rates of the continuous phase $Q_c$ could be reached with couple 2 corresponding to a less resistive geometry. Consequently, for fixed $Q_d$, another dripping-jetting transition could be observed by increasing $Q_c$ beyond the range shown in Fig.~\ref{Fig:WO30mu}. Fig.~\ref{Fig:DripJet} shows for $Q_d=50~\mu$L/min that this transition occurs for $Q_c$ at around $5200\pm 500$ $\mu$L/min. This value is in good agreement with the prediction of the numerical model as described below.

\section{Modelling}
\label{Modelling}
With the goal to predict the droplet radius generated in the dripping regime using the Raydrop, we propose in this section to model the non-embedded {\it co-flow-focusing} configuration in transient and then using the quasi-static approach. After a validation with the experimental results, we show how to determine the droplet size in the dripping regime, as well as the dripping to jetting transition, when varying the continuous flow rate alone.

\subsection{Transient}
\label{Transient}
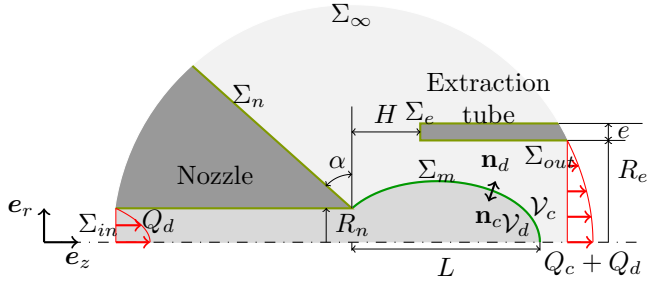
\begin{figure}[h!]
	\centering
	\definecolor{mycolor1}{rgb}{0.0,.6,0.0}
\definecolor{mycolor2}{rgb}{0.5,.6,0.0}

\begin{tikzpicture}[scale=0.45]

\draw[fill=gray!10!white,gray!10!white] (7,0) arc (0:180:7);  
\draw[fill=gray!30!white,gray!30!white] (0,0) -- (-7,0)--(-7,1)--(-3.5,1)--(0,1)--cycle;  
\draw[white, fill=white] plot  (-6.9,0) rectangle (-7.5,1); 
\draw[white, fill=white] plot  (6.3,0) rectangle (9,3); 

\draw[gray!30!white,thick,fill=gray!30!white]  plot[domain = 0:145.5, samples = 40, variable = \i]  ({2.5+3*cos \i}, {1.8*sin \i}) --(0,0); 
\draw[mycolor1,thick] plot [domain = 0:145.5, samples = 40, variable = \i]  ({2.5+3*cos \i}, {1.8*sin \i}); 
\node at (2.5,2.1) {$\Sigma_{m}$};
\node at (4.8,0.5) {$\V_d$};
\node at (5.6,1.0) {$\V_c$};

\draw[mycolor2,fill=gray!80!white,thick] (-7,1) -- (0,1) -- (-5,5.5) -- (-6,5.); \node at (-3.,4.3) {$\SigmaN$};
\draw[mycolor2,fill=gray!80!white,thick] (2,3.5) -- (2,3) -- (7,3) -- (7,3.5)--cycle; 
\node at (2,3.8) {$\Sigma_e$}; 
\draw[white,fill=white] (7,0) arc (0:180:7) -- (-8,0) arc (180:0:8) -- cycle;
\draw[dashdot] (-8.5,0) -- (8.5,0);

\draw[ultra thin,<->] (0,3.25)--(2,3.25) node[midway,above] {$H$};
\draw[ultra thin] (0,1) -- (0,4);
\draw[ultra thin,<->] plot [domain = 90:139, samples = 40, variable = \i]  ({1*cos \i}, {1+1*sin \i}); \node[above] at ({cos(115)},{1+sin(115)}) {$\alpha$}; 
%
%


\draw[ultra thin,->] (-0.75,0) -- (-.75,1) node [midway,right] {$R_n$};


\draw[ultra thin] (8.5,3.5) -- (6.05,3.5);
\draw[ultra thin] (8.0,3) -- (6.3,3);
\draw[ultra thin,->] (7.5,0) -- (7.5,3) node [pos=.7,right] {$R_{e}$};
\draw[ultra thin,<->] (7.5,3) -- (7.5,3.5) node [pos=.5,right] {$e$};


\draw[ultra thin] (0,0) -- (0,-.25);
\draw[ultra thin] (5.5,0) -- (5.5,-.25);
\draw[ultra thin,<->] (0,-.2) -- (5.5,-.2) node [pos=.5,below] {$\disp$};

\draw[red] plot [domain = 0:1, samples = 40, variable = \i]  ({-6.9+1-\i*\i}, {\i}); 
\draw[red] plot [domain = 0:1, samples = 40, variable = \i]  ({-6.9}, {\i}); 
\draw[red,thick,->] (-6.9,0)--({-6.9+1},0);
\draw[red,thick,->] (-6.9,0.5)--({-6.9+1-.25},0.5); \node[above] at ({-6.9+1-.25+.5},0.) {$Q_d$}; 

\draw[red] plot [domain = 0:1, samples = 40, variable = \i]  ({6.3+.75-.75*\i*\i}, {3*\i}); 
\draw[red] plot [domain = 0:1, samples = 40, variable = \i]  ({6.3}, {3*\i}); 
\draw[red,thick,->] ({6.3}, {3*0})--({6.3+.75-.75*0*0}, {3*0});
\draw[red,thick,->] ({6.3}, {3*0.25})--({6.3+.75-.75*0.25*0.25}, {3*0.25});
\draw[red,thick,->] ({6.3}, {3*0.50})--({6.3+.75-.75*0.50*0.50}, {3*0.50});
\draw[red,thick,->] ({6.3}, {3*0.75})--({6.3+.75-.75*0.75*0.75}, {3*0.75});  \node[below] at ({6.3+.75-.75*0*0}, {3*0}) {$Q_c+Q_d$}; 

\node at (0,6.7) {$\Sigma_{\infty}$};
\node at (5.8,2.5) {$\Sigma_{out}$};
\node at (-7.4,.5) {$\Sigma_{in}$};

\draw[black,thick,->] (-9,0) -- (-8,0) node [below] {$\boldsymbol{e}_z$};  
\draw[black,thick,->] (-9,0) -- (-9,1) node [left] {$\boldsymbol{e}_r$};

\node at (-4,2) {Nozzle};
\node[align=center] at (4,4.25) {Extraction \\ tube};  

 \draw[->,thick] (4.1,1.5) --++ (+0.1,+.3) node[anchor=south] {$\vn_d$}; 
 \draw[->,thick] (4.1,1.5) --++ (-0.1,-.3) node[anchor=north] {$\vn_c$}; 

\end{tikzpicture}
	\caption{Sketch of the droplet generation inside a Raydrop.}
	\label{Sketch}
\end{figure}

Geometry of the non-embedded {\it co-flow-focusing} configuration is considered in an axisymmetric coordinate system $(r,z)$, with $r$ and $z$ the radial and axial coordinates, respectively, as sketched in \figref{Sketch}. 
The properties and parameters for the two phases will be given with the generic subscript $i$ referring to $d$ and $c$ for the dispersed and the continuous phases, respectively. Each phase has thus a density $\rho_i$, a dynamic viscosity $\mu_i$ and a flow rate $Q_i$. The time-dependent volumes of the two phases are denoted by $\V_i$ and are separated by the surface $\Sigma_m$, characterized by an interfacial tension $\gamma$. The surface of the nozzle and the extraction capillary are denoted by $\Sigma_n$ and $\Sigma_e$, respectively. The domain considered is truncated at $\Sigma_{\infty}$, $\Sigma_{\rm{in}}$ and $\Sigma_{\rm{out}}$, which are sufficiently far from the droplet generation zone to not affect the results. The dispersed flow rate $Q_d$ is injected in the form of a fully developed Poiseuille flow through the capillary of the nozzle at $\Sigma_{\rm{in}}$, and both fluids leave the device through the cross-section of the extraction capillary $\Sigma_{\rm{out}}$ at a flow rate $Q_c+Q_d$. In addition to the geometrical parameters already defined in \tabref{tbl:data}, one can define $\disp$ as the distance taken on the symmetry axis from the nozzle tip to the tip of the meniscus, as sketched in \figref{Sketch}.

The motion of the fluids is governed by the continuity and the Navier--Stokes equations, 
\begin{alignat}{7}
\label{NS}
\vnabla \cdot \vv_i = 0 \,, \quad \rho_i (\partial_t \vv_i + (\vv_i \cdot \vnabla) \vv_i) = \vnabla \cdot \stress_i  \,, \quad \mbox{at } \V_i \,, 
\end{alignat}
where $\vv_i$ is the velocity vectors and $\stress_i=-p_i \id + \mu_i \left[ \vnabla \vv_i + \left( \vnabla \vv_i \right)^T \right]$ is the stress tensor of the phase $i$ with $p_i$ being the pressure field and $\id$ the identity matrix. Defining $\vn_i$ as the outer normal to the domain $\V_i$ at any of its boundaries, as sketched in \figref{Sketch}, the no-stress boundary condition at $\Sigma_{\infty}$ write
\begin{align}
\vn_c \cdot \stress_c = \vzero \qquad \mbox{at } \Sigma_{\infty}  \,.
\end{align}
The flow being considered as fully developed (i.e. parallel flow), upstream of the nozzle as well as downstream of the extraction capillary, 
\begin{subequations}
\begin{alignat}{2}
 \vn_d \cdot \stress_d &= P_d \,\vn_d \qquad \mbox{at } \Sigma_{\rm{in}}  \,, \\
 \vn_c \cdot \stress_c &= P_c \, \vn_c \, \qquad \mbox{at } \Sigma_{\rm{out}}  \,,
\end{alignat}
\end{subequations}
where the pressures, $P_d$ and $P_c$, correspond to those ones imposing the given flow rates 
\begin{subequations}
\label{Qo}
\begin{align}
\int_{\Sigma_{\rm{out}}}  \ve_z \cdot \vv_c \, \dd \Sigma &= Q_d+ Q_c \,,  \\
{\V}_d(t) - {\V}_d(0) &= Q_d \, t,
\label{eq1.4b}
\end{align}
\end{subequations}
with $t$ the time.
No-slip boundary conditions at the nozzle and at the extraction capillary are applied
\begin{subequations}
\begin{align}
 \vv_c &= 0  \qquad \mbox{at } \Sigma_{n} \cup \Sigma_{e}  \,,  \\
 \vv_d &= 0  \qquad \mbox{at } \Sigma_{n}  \,.
\end{align}
\end{subequations}
The continuity of velocity and stress balance at the interface, which is pinned at the nozzle tip, writes
\begin{subequations}
\begin{alignat}{3}
\vv_d  &= \vv_c \, &\qquad& \mbox{at } \Sigma_{m}  \,, \\
\vn_d \cdot \stress_d + \vn_c \cdot  \stress_c &= \vnabla_S \cdot \left( \gamma \id_S \right) \, &\qquad& \mbox{at } \Sigma_{m}  \,,
\end{alignat}
\end{subequations}
where $\id_S= \id - \vn_i \vn_i$ is the surface identity tensor. The kinematic condition writes
 \begin{align}
\label{kinematic}
  \vv_d  \cdot \vn_d  &=   \partial_t \vx \cdot \vn_d \, \qquad \mbox{at } \Sigma_{m}  \,, 
\end{align}
where $\vx \in \Sigma_{m}$ and $ \partial_t \vx \cdot \vn_d$ represents the normal velocity of the interface.

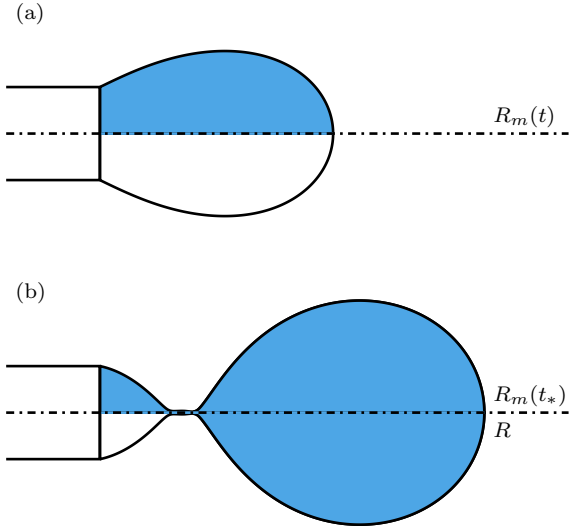
\begin{figure}
	\centering

\definecolor{mycolor1}{rgb}{0.3,0.65,0.9}
\definecolor{mycolor2}{rgb}{0.3,0.65,0.9}
\definecolor{mycolor3}{rgb}{1,1,1}

\begin{tikzpicture}[baseline]



\begin{axis}[
scale only axis,
width={.4\textwidth},
height={.2\textwidth},
xmin=-2,
xmax=10,
ymin=-3,
ymax=3,
hide axis,
name=unouno,
]

\addplot[black,line width=1,fill=mycolor1,area legend] table[x expr=\thisrowno{0},y expr=\thisrowno{1}*+1] {./SketchContour3.txt};
\addplot[black,line width=1,fill=white      ,area legend] table[x expr=\thisrowno{0},y expr=\thisrowno{1}*-1] {./SketchContour3.txt};


\draw[mycolor2,line width=1] (axis cs: 0,0)--(axis cs: 4.976260194304993,0);
\draw[black,line width=1,dashdotted] (axis cs: -2,0)--(axis cs: 10,0);
\draw[black,line width=1] (axis cs: -2,1)--(axis cs: 0,1)--(axis cs: 0,-1)--(axis cs: -2,-1);

\node[black,anchor=west,xshift=0.0cm,yshift=0.0cm] at (rel axis cs:.85,.56) {$\Rmen(t)$};

\node[black,anchor=north west,xshift=0.0cm,yshift=0.0cm] at (rel axis cs:0,1) {(a)};

\end{axis}

\begin{axis}[
yshift={-.2\textwidth},
scale only axis,
width={.4\textwidth},
height={.2\textwidth},
xmin=-2,
xmax=10,
ymin=-3,
ymax=3,
hide axis,
]

\addplot[black,line width=1,fill=mycolor3] table[x expr=\thisrowno{0},y expr=\thisrowno{1}*-1] {./SketchContour.txt};
\addplot[black,line width=1,fill=mycolor2] table[x expr=\thisrowno{0},y expr=\thisrowno{1}*-1] {./SketchContour2.txt};

\addplot[black,line width=1,fill=mycolor2] table[x expr=\thisrowno{0},y expr=\thisrowno{1}*+1] {./SketchContour.txt};
\addplot[black,line width=1,fill=mycolor2] table[x expr=\thisrowno{0},y expr=\thisrowno{1}*+1] {./SketchContour2.txt};


\draw[mycolor1,line width=1] (axis cs: 0,0)--(axis cs: 8.202387545197386,0);

\draw[black,line width=1,dashdotted] (axis cs: -2,0)--(axis cs: 10,0);
\draw[black,line width=1] (axis cs: -2,1)--(axis cs: 0,1)--(axis cs: 0,-1)--(axis cs: -2,-1);

\node[black,anchor=west,xshift=0.0cm,yshift=0.0cm] at (rel axis cs:.85,.56) {$\Rmen(t_*)$};
\node[black,anchor=west,xshift=0.0cm,yshift=0.0cm] at (rel axis cs:.85,.44) {$R$};

\node[black,anchor=north west,xshift=0.0cm,yshift=0.0cm] at (rel axis cs:0,1) {(b)};
\end{axis}


\end{tikzpicture}
	\caption{Sketch of the instantaneous equivalent radius of the meniscus $\Rmen(t)$, and of the drop $R$, (a) for an arbitrary time $t$ and (b) for the pinch-off time $t_*$.}
	\label{Sketch20}
\end{figure}

The instantaneous equivalent radius of the meniscus $\Rmen(t)$ is defined as the radius of a sphere of a volume equivalent to the one of the dispersed phase beyond the nozzle and bounded by the meniscus surface $\Sigma_m$, namely
\begin{equation}
\frac43 \pi \Rmen^3 (t) =   \frac12 \int_{\Sigma_m(t)} r \,\ve_r \cdot \vn_d \, \dd \Sigma \,.
\end{equation}
It should be noted that at the pinch-off time $t_*$, the volume preceding the neck is the one corresponding to the droplet $\frac43 \pi R^3$. In \figref{Sketch20}, the volumes of revolution reconstructed from the shaded areas are related to $\Rmen$ and $R$. 
The meniscus shape and volume for an arbitrary time and for the pinch-off time are sketched in \figref{Sketch20}(a) and (b), respectively. The instantaneous equivalent radius of the meniscus $\Rmen(t)$ is time-dependent and shadowed at both times, whereas the radius of the drop $R$ is only defined at the pinch-off time $t_*$, and is therefore not time-dependent.

The equations are adimensionalised using $R_n$ for the length scale, $\gamma / \mu_c$ for the velocity scale and $ \gamma/R_n$ for the pressure scale, leading to the following dimensionless flow parameters 
\begin{eqnarray}
 &\displaystyle \Ca_c = \frac{\mu_c Q_c}{\gamma \pi R_{e}^2}\,, \quad  \bar R_e= \frac{R_e}{R_n}\,, \quad  \bar Q= \frac{Q_d}{Q_c}\,,&  \label{Ca}  \\
 & \displaystyle  \lambda = \frac{\mu_d}{\mu_c}\,, \quad \La = \frac{\rho_c \gamma R_n}{\mu_c^2} \quad \text{and} \quad \phi = \frac{\rho_d}{\rho_c}\,,& \label{La}
\end{eqnarray}
with $\Ca_c$ the capillary number for the continuous phase, $\bar R_e$ the radius ratio, $\bar Q$ the flow rate ratio, $\lambda$ the viscosity ratio, $\La$ the Laplace number and $\phi$ the density ratio. The capillary number for the dispersed phase is therefore $\Ca_d = \bar Q \, \Ca_c$. Finally, the other geometrical parameters are all made dimensionless relative to the diameter of the nozzle tip:
\begin{equation}
\bar H = \frac{H}{R_n}\,, \quad \bar e=\frac{e}{R_n}\,, \quad \bRmen = \frac{\Rmen}{R_n}\,, \quad \bar L = \frac{L}{R_n} \,.
\end{equation}

The system of equations \eqref{NS}-\eqref{kinematic} governs the droplet formation using the Raydrop. The domain variables are $p_i$ and $\vv_i$, the surface variable is $\vx$, and the global variables are $P_i$. This system of equations is solved using the finite element method (FEM) with the help of the software Comsol and quadratic Lagrangian elements, with exception of the pressure for which linear elements have been used. For the deformable domain, the \emph{moving mesh} application mode has been used.  It implements the Arbitrary Lagrangian-Eulerian (ALE) method combined with the Boundary Arbitrary Lagrangian-Eulerian (BALE) method previously developed in~\cite{Rivero18}. In every simulation, the initial shape of the meniscus is half a sphere out of the nozzle. Transient simulations have been carried out using a first-order backward-Euler time discretization until the pinch-off time. The criteria for the pinch-off is that the neck radius is $2$\% of the nozzle tip radius. 

Finally, we note that most of the results presented below have been obtained with the Stokes equations, i.e. with the left-hand side of the momentum balance in eqn~(\ref{NS}) set to $\bf 0$. We have indeed checked numerically (not shown) that for the maximum Reynolds number corresponding to our experimental conditions (see \tabref{tbl:data}), or equivalently for a maximum Laplace number $\La$ of about 100, inertia has no influence on these results. Therefore, all results presented below are for $\La=0$, unless specified otherwise.

\subsection{Quasi-static approach}
\label{Quasi-static approach}
In general, the system behavior is transient but considering the quasi-static (QS) limit is of great interest. This limit corresponds to the situation of $Q_d \rightarrow 0$, meaning that the flow of the dispersed phase is negligible, and thus set to zero, i.e. $Q_d=0$. Consequently,  the system of equations \eqref{NS}-\eqref{kinematic} can be solved for stationary solutions, cancelling the time-derivatives in eqn~\eqref{NS} and \eqref{kinematic}, and parametrizing the volume of the liquid meniscus with its quasi-static equivalent radius $\bRmen$, instead of using time $t$ in eqn~\eqref{eq1.4b}. This QS approach thus facilitates the parametrical analysis and allows to provide an explanation to the underlying mechanism behind the drop formation. By convenience, quasi-static simulations have been carried out using parametric continuation in $\bdisp$ instead of $\Ca_c$, this latter being considered as a simulation output.

For comparison purposes, we show in Table~\ref{Equivalent radius} the correspondence between the equivalent radii obtained from the transient and the quasi-static approaches.
\begin{table}
\centering
\begin{tabular}{ll}
\multicolumn{2}{l}{\underline{Transient ($\Ca_d >0$):}} \\
Equivalent radius of the meniscus  & $\bRmen(t)$     \\
$\quad ...$ at pinch-off time $t_*$ & $\bRmen(t_*)$  \\
$\quad \quad ...$ preceding the pinch-off location & $\bR$   \\
$\qquad \quad ...$ in the QS limit for $\Ca_d \rightarrow 0$  & $\bR^{\rm QS}$ \\
\multicolumn{2}{l}{\underline{Quasi-static ($\Ca_d = 0$):}} \\
Equivalent radius of the meniscus  & $\bRmen$ \\
$\quad ...$ at the folding bifurcation point & $\bRmen^D$     \\
\end{tabular}
\caption{Dimensionless equivalent radii obtained in transient, as defined in \figref{Sketch20}, and with the quasi-static approach.}
\label{Equivalent radius}
\end{table}

\subsection{Model validation}
\label{Model validation}
\begin{figure*}[h!]
	\centering
	\includegraphics[width=\textwidth]{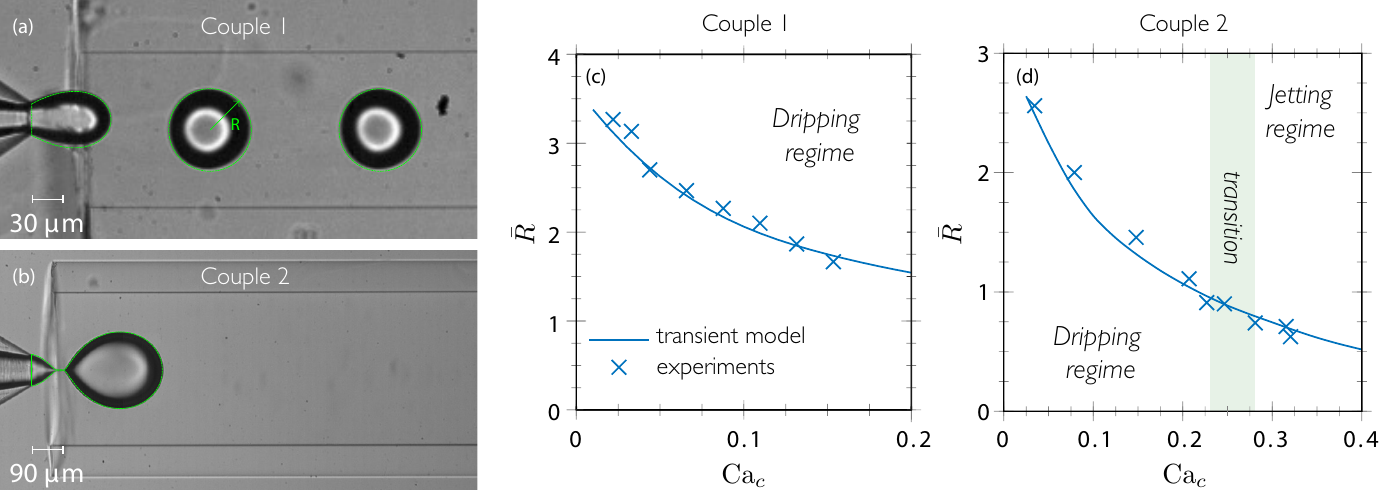}
	\caption{Comparison between transient simulations (green lines) and experimental pictures for the formation of water drops in mineral oil with properties reported in Table~\ref{tbl:data}: (a) meniscus for $Q_c=150\,\rm{\mu L/min}$ ($\Ca_c=0.066$) and $Q_d=25\,\rm{\mu L/min}$ ($\Ca_d=0.01$); (b) meniscus at pinch-off for $Q_c=700\,\rm{\mu L/min}$ ($\Ca_c=0.034$) and $Q_d=20\,\rm{\mu L/min}$ ($\Ca_d=0.001$); (c-d) comparisons of droplet size for various values of $\Ca_c$ in the dripping regime for (c) and showing for (d) the experimental dripping-jetting transition (shaded area), as identified in \figref{Fig:DripJet}}
	\label{Compara0}
\end{figure*}

The model in the Stokes limit computed first in transient is validated by comparison with experimental data, as shown in \figref{Compara0}. 
\figref{Compara0}(a) displays the superposition of the droplet shape with a circle of radius $R$ (green line) obtained from the equivalent radius of the meniscus $R_m$ predicted by the model for couple 1. In \figref{Compara0}(b), the interfacial shape of a drop in formation is shown just before the pinch-off of the meniscus for couple 2. In both cases, an excellent agreement is observed. In \figref{Compara0}(c-d), the experimental and numerical drop sizes are compared for a wide range of $\Ca_c$, showing again an excellent agreement for the two geometries (couples 1 and 2). It can be observed that as the continuous flow rate increases, the generated drop becomes smaller. The shaded area in \figref{Compara0}(d) corresponds to the dripping-jetting transition identified experimentally in \figref{Fig:DripJet} in the range $0.23 \lesssim \Ca_d \lesssim 0.28$. Finally, note that $Q_d$ has been kept small enough to ensure $\bar Q \ll 1$, such as it should have no influence on the droplet size, then suggesting that the QS limit is applicable.

\begin{figure}[h!]
	\centering
	\includegraphics[width=0.45\textwidth]{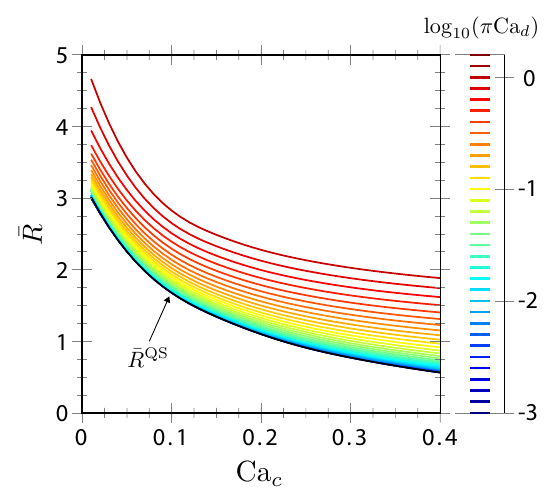}
	\caption{Influence of the $\Ca_d$ on the generated drop radius as a function of $\Ca_c$ computed in transient considering couple 1 (see Table~\ref{tbl:data}); $\bar R^{\rm QS}$ in black line corresponds to $\Ca_d = 4\times 10^{-6}\pi^{-1} $, i.e. the quasi-static limit.}
	\label{Qi}
\end{figure}

In order to verify the range of validity of the QS limit, \figref{Qi} shows the predictions of the droplet size in transient for various values of $\Ca_d$, as well as in the QS limit as represented by the black solid line, denoted as $\bR^{\rm QS}$ and effectively obtained for $\Ca_d = 4\times 10^{-6}\pi^{-1}$.
This limit lies slightly below the numerical curve plotted in \figref{Compara0}(c), as represented by the orange area in \figref{Qi} for $\log_{10}(\pi \Ca_d) \approx -0.67$. \figref{Qi} shows that for increasing values of $\Ca_d$, the system leads to a monotonous  (yet logarithmic) increase of the drop size, as compared to the QS limit prediction. 

\begin{figure}[h!]
	\centering
	\includegraphics[width=0.45\textwidth]{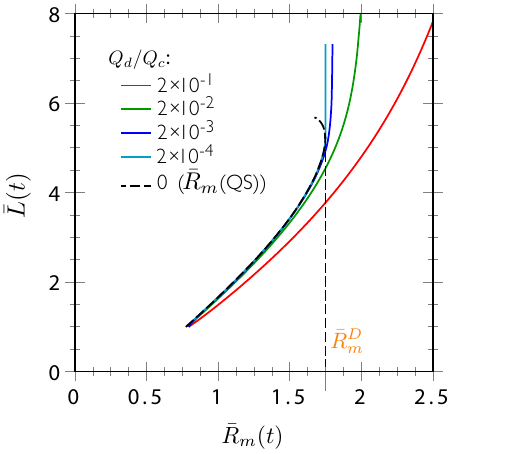}
	\caption{Transient evolution of the meniscus length $\bdisp$ as a function of the meniscus equivalent radius $\bRmen$ for different values of $\bar Q=Q_d/Q_c$ and $\Ca_c=0.1$, considering couple 1 (see Table~\ref{tbl:data}). The equivalent radius of the meniscus calculated with quasi-static simulations is plotted in black dashed line.}  
	\label{Compara2}
\end{figure}

In order to understand how the flow rate of the continuous phase affects the size of the drop as a function of the geometry and of the properties of the given pair of fluids, the comparison between the transient and the quasi-static solutions is further investigated. In \figref{Compara2}, the dynamics of the meniscus shape predicted from a quasi-static simulation ($\bar Q = 0$), as well as from transient simulations of several values of $\bar Q$ is presented for a given value of $\Ca_c$.
Parameters $\bdisp$ and $\bRmen$ were chosen as representative quantities of the meniscus shape and are plotted as trajectory lines following the time during the formation process. One notices that the lines diverge as $\bar Q \rightarrow 0$. Remarkably, solving the model using the quasi-static simulation for exactly $\bar Q = 0$ depicts the existence of a turning-point corresponding to a folding bifurcation. This is analogous to the detachment of a pendant drop as discussed in the introduction and further developed below. We note $\bRmen^D$ the value of $\bRmen$ at the turning-point, as labeled in Table~\ref{Equivalent radius}. 

In the transient simulations with finite values of $\bar Q$, it can be observed (i) that the meniscus is less elongated for larger $\bar Q$ and given $\bRmen$, especially close to the turning-point, which naturally leads to larger droplets as shown in \figref{Qi}, and (ii) that larger menisci, i.e. $\bRmen>\bRmen^D$, can be attained for finite dispersed flow rates. At that stage, $\bdisp$ exhibits a large increase for a small increase of $\bRmen$, due to the elongation concentrated in the neck region shown in \figref{Compara0}(b), i.e. the menisci evolve towards the pinch-off, leading to the formation of a drop of size $\bR$. 

\begin{figure}
	\centering
	\includegraphics[width=0.48\textwidth]{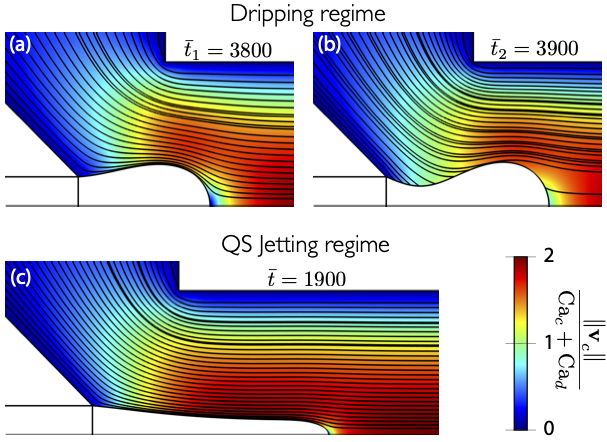}
	\caption{Axisymmetric simulations in transient using couple 1 (see Table~\ref{tbl:data}) and $\Ca_d=8 \cdot 10^{-5} $ (QS limit). The flow field is represented by streamlines and rescaled velocity norm $||\vv_c||/(\Ca_c+\Ca_d)$ as color code. The timescale is $R_n\mu_c/\gamma=7\,\mu$s. (a--b) Dripping regime with $\Ca_c=0.1$ for a time before (a) and after (b) the folding bifurcation; (c) QS-jetting regime for $\Ca_c=0.3>\Ca_c^{*}$ (see section~\ref{Inviscid droplets} for details). }  
	\label{Streamlines_drip}
\end{figure}

In \figref{Streamlines_drip}, the flow field and meniscus shape are represented at two different times. While in \figref{Streamlines_drip}(a) the volume exhibits a quasi-steady meniscus since $\bRmen(\bar{t}_1)<\bRmen^D$, in \figref{Streamlines_drip}(b) the volume is no longer in equilibrium since $\bRmen (\bar{t}_2) > \bRmen^D$, leading to the formation of a neck.
On the one hand, it can be observed that, for a quasi-static meniscus, streamlines are tangent to the meniscus revealing the quasi-static character of the flow, i.e. the right-hand side term of eqn\eqref{kinematic} is negligible and thus set to zero, i.e. $Q_d=0$. On the other hand, for larger volumes, the quasi-static meniscus no longer exists and it exhibits a neck which eventually breaks for later times. In this situation, the meniscus evolves dynamically as revealed by the streamlines which are no longer tangent to the meniscus, and the right-hand side of eqn~\eqref{kinematic} is not negligible anymore, even in the limit of $\bar Q \rightarrow 0$. Finally, the transient solution shown in \figref{Streamlines_drip}(c) for a larger $\Ca_c$ depicts the shape of a jet with streamlines again parallel to the interface, thus suggesting a quasi-static solution like in \figref{Streamlines_drip}(a). As it will be shown below, this solution corresponds to the jetting regime in the quasi-static limit, occurring at a $\Ca_c$ larger than the transition value, denoted by $\Ca_c^*$.

\section{Parametric analysis}
\label{Parametric analysis}
Results obtained so far support the quasi-static approach for two reasons: (i) theoretically, the occurrence in \figref{Compara2} of a folding bifurcation in this limit allows for transition tracking between dripping and jetting modes; (ii) experimentally, the size of the droplets shown in \figref{Fig:WO30mu} barely varies with $Q_d$ at constant $Q_c$, suggesting that size predictions in the limit of $Q_d \rightarrow 0$ should also apply for finite $Q_d$, which then only influences the frequency of droplet generation, at least to some extend. In this section, quasi-static simulations are therefore used as a predictive tool to evaluate the influence of fluid properties and geometric parameters on the drop formation.

\subsection{Inviscid droplets}
\label{Inviscid droplets}
\begin{figure}[h!]
\centering
\includegraphics[width=0.4\textwidth]{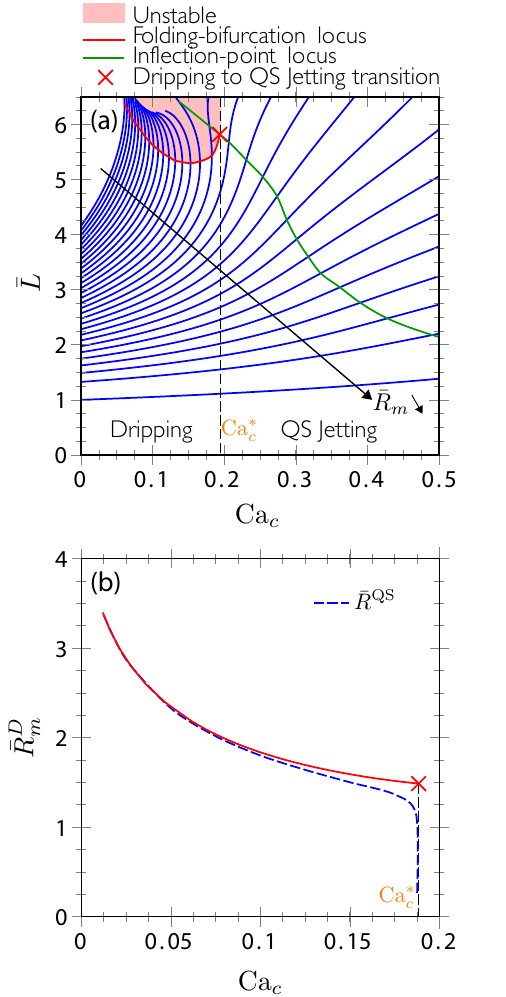}
\caption{Dripping mode obtained with quasi-static simulations for the geometry of couple 1 and in the inviscid limit $\lambda=0$: influence of (a) $\bRmen$ and $\Ca_c$ on $\bdisp$ and (b) $\Ca_c$ on $\bRmen^D$. The blue dashed line represents the equivalent radius $\bR^{\rm QS}$ as defined in Table~\ref{Equivalent radius}, while the red cross indicates the dripping to jetting transition at $\Ca_c^*$.}  
\label{Mechanism}
\end{figure}

Starting with the inviscid droplet limit, i.e. $\lambda=0$, we plot in \figref{Mechanism}(a) the quasi-static meniscus length $\bdisp$ as a function of $\Ca_c$ for constant equivalent radii of the meniscus $\bRmen$ (blue lines), the arrows indicating the decrease of $\bRmen$. 
It can be observed that in some range of $\Ca_c$ corresponding to large values of $\bRmen$, the curves of constant $\bRmen$ exhibit a turning-point. The loci of the turning-points are then depicted by the red curve, as determined by $\partial \Ca_c / \partial  \bdisp \rvert_{\bRmen} = 0 $.  The shadowed region above this curve is therefore the unstable region. Consequently, a dripping occurs at a fixed $\Ca_c$ as the volume of the meniscus is quasi-statically increased up to the red curve corresponding to $\bRmen=\bRmen^D$. Quasi-static solution no longer exists for higher volumes, as already shown in \figref{Compara2} for $\Ca_c=0.1$ and $\lambda=1/23$.   

Notably, the locus of turning points stops at $\Ca_c^*$ (red cross in \figref{Mechanism}(a)), thus representing the maximum $\Ca_c$ for which the dripping regime in the QS limit occurs. Indeed, as a matter of fact, this maximum no longer exists on the right of the locus of the inflection-point (green curve), determined by $ \partial^2 \Ca_c / \partial  \bdisp^2 \rvert_{\bRmen} = 0 $. Quasi-static and transient simulations show that for $\Ca_c>\Ca_c^*$, a jet is established as was shown in \figref{Streamlines_drip}. In \figref{Mechanism}(a), and for a fixed $\Ca_c > \Ca_c^*$, as the volume of the meniscus increases, the tip of the jet advances and the jet is formed behind. Remarkably, the form of the jet does not vary behind the tip, which recalls the steady shape of the fluid cone in the tip streaming process~\cite{Suryo06,Evangelio16}.

In \figref{Mechanism}(b), the folding bifurcation locus is shown in the ($\bRmen^D$,$\Ca_c$)-plane. It can be observed that the equivalent radius is smaller for larger values of the continuous flow rate, as the viscous forces exerted by the continuous phase on the meniscus is larger, inducing the detachment of a smaller droplet. This recalls the analogy with the pendant droplet, provided the hydrodynamic forces are substituted to the gravity forces. We also plot in blue dashed line the equivalent radius $\bR^{\rm QS}$ corresponding to the volume of revolution preceding the neck, as obtained in transient in the quasi-static limit, i.e. for $\Ca_c \rightarrow 0$ (see Table~\ref{Equivalent radius}).  As $\Ca_c$ increases, the position of the neck moves from the nozzle at $\Ca_c \rightarrow 0$ to the tip of the meniscus at $\Ca_c^*$, for which the neck disappears. This explains the dramatic decrease of $\bR^{\rm QS}$ when approaching the jetting transition, even though it only occurs for inviscid droplets and is therefore not really physical. Indeed, a finite viscosity of the dispersed phase is found to regularise this singularity (see \figref{Qi}).

Now, for lower $\Ca_c$, $\bR^{\rm QS}$ is shown to follow the same behavior than $\bRmen^D$, even though it remains smaller because of the volume that remains attached to the nozzle behind the neck. Consequently, it confirms that except near the dripping-jetting transition, the value $\bRmen^D$ is well representative of the drop radius determined in transient in the quasi-static limit, but without relying on the pinch-off formation.

Finally, it is worth mentioning that the equivalent radius of the meniscus for dripping in \figref{Mechanism}(b) can be correlated by 
\begin{equation}
\bRmen^D \approx 0.84 \,\Ca_c^{-0.34}\qquad (\lambda=0). \label{correlCa0}
\end{equation}
The prefactor and mostly the exponent are similar to those found by Lan {\it et al.}~\cite{Lan15} for dripping in the a co-flow configuration, namely $\bR=1.07 \, \Ca^{-0.34}Re_d^{0.06}$ with $Re_d=2\rho_dQ_d/(\pi\mu_d R_n)$ the Reynolds number of the dispersed flow. Apart from the slight influence of inertia, i.e. $Re_d^{0.06}\approx1.2$, not included in eqn~(\ref{correlCa0}) since obtained for a Stokes flow and $\lambda=0$, it shows that the drop size predictions in the dripping mode are comparable in co-flow and {\it co-flow-focusing}.

\subsection{Viscous droplets}
\label{Viscous droplets}
\begin{figure}[h!]
	\centering
	\includegraphics[width=0.45\textwidth]{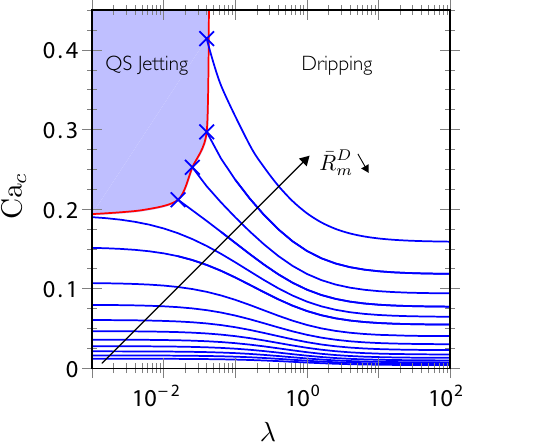}
	\caption{Influence of $\lambda$ in the quasi-static dripping regime on the value of $\Ca_c$ necessary to produce a drop from a meniscus equivalent radius $\bRmen^D$ (blue lines). The red curve corresponds to the dripping-jetting transition for $\Ca_c^*$. The geometry of couple 1 has been used.}  
	\label{ViscRat}
\end{figure}

Next, the influence of the viscosity ratio $\lambda$ on the meniscus equivalent radius is considered. Each blue line in \figref{ViscRat} corresponds to a constant meniscus equivalent radius $\bRmen^D$ in the ($\lambda$,$\Ca_c$)-plane, while the black arrow indicates the decrease of the meniscus volume.
We observe that, as $\lambda$ increases, smaller values of $\Ca_c$ are needed to generate a drop of the same size. This can be explained by the fact that for inviscid droplet, only the normal component of the viscous stresses exerted at the interface contributes to the net force responsible for the droplet detachment, while, as $\lambda$ increases, the tangential (or shear) component is added to the force, provoking the detachment of smaller droplets. For the same reason, the dripping-jetting transition occurs for larger values of $\Ca_c^*$ as $\lambda$ increases, as shown by the red curve in \figref{ViscRat}, which starts from the value of $\Ca_c^*=0.188$ in the limit of inviscid droplets, to large and possibly asymptotically infinite value for viscous droplets corresponding to $\lambda \gtrsim 5\times 10^{-2}$. It thus appears that beyond this value, the viscous shear stress exerted on the droplet is always large enough to ensure the dripping mode.

\begin{figure}[h!]
	\centering
	\includegraphics[width=0.4\textwidth]{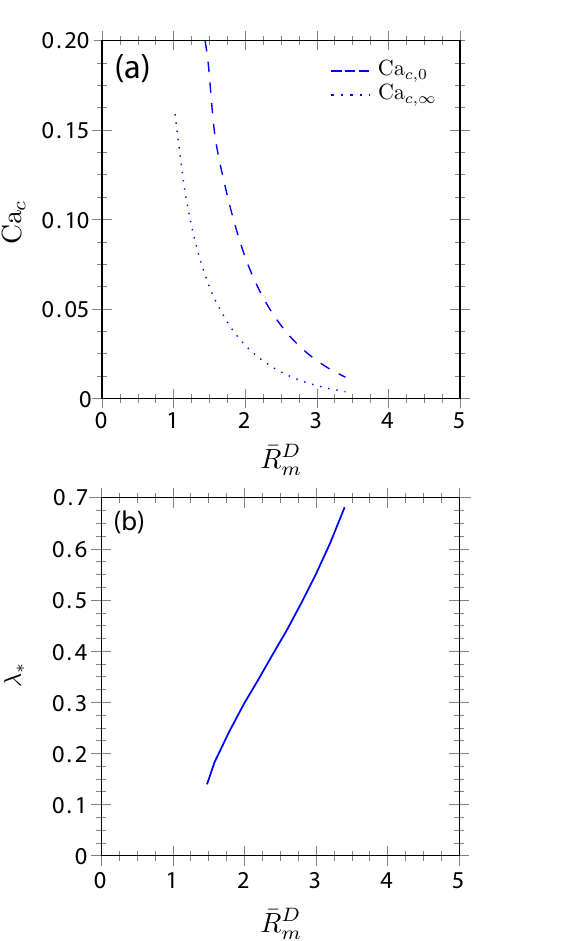}
	\caption{Computed functions involved in eqn~(\ref{correlation}), which describes the influence of $\lambda$ on $\Ca_c$ in the quasi-static dripping regime for the geometry of couple 1: (a) $\Ca_{c,0}$, $\Ca_{c,\infty}$ and (b) $\lambda_* $. Note that the $\Ca_{c,0}$ (dashed line) is identical to the red curve in \figref{Mechanism}.}  
	\label{ViscRat1}
\end{figure}

Practically, the influence of $\lambda$ on $\Ca_c$ for the dripping mode can be described by a correlation involving the limits for $\lambda\rightarrow 0$ and $\lambda \rightarrow \infty$,
\begin{align}
\Ca_c (R_m^D,\lambda)=\frac{\lambda_* (\bRmen^D) \,\Ca_{c,0} (\bRmen^D) + \lambda \,\Ca_{c,\infty} (\bRmen^D) }{\lambda_* (\bRmen^D) + \lambda} \,, \label{correlation}
\end{align}
where $\Ca_{c,0}$ and $\Ca_{c,\infty}$ are the $\Ca_c$ numbers that produce menisci of size $\bRmen^D$ in both limits, respectively, whereas the transition is centered at $\lambda_*$. These functions are plotted in \figref{ViscRat1}(a-b),
the correlations of which are $\Ca_{c,0}\approx 0.5952 (\bRmen^D)^{-2.936}$ (as reverted from eqn~(\ref{correlCa0})), $\Ca_{c,\infty}\approx 0.1666 (\bRmen^D)^{-2.496}$ and $\lambda_* \approx 0.09497(\bRmen^D)^{1.609}$, applicable in the range $1.5 \lesssim \bRmen^D \lesssim 3.5$. Using these correlations in eqn~(\ref{correlation}) allows to estimate the dimensionless droplet size $\bRmen^D$ for a given viscosity ratio $\lambda$ and a given capillary number of the continuous phase $\Ca_c$, provided the geometry verifies the dimensionless parameters corresponding to couple 1, namely $\bar R_e=5$, $\bar H=3.66$, $\bar e=5$ and $\alpha = 50^\circ$. The sensitivity of the droplet size on these parameters is analysed in the next section.

Figure~\ref{ViscRat1}(a) shows that droplets generated in the infinite viscous limit are always smaller than those generated in the inviscid limit, for a fixed $\Ca_c$, because of the additional shear stresses contributing to the force involved in the dripping mechanism. This argument should also be completed by the fact that if the viscosity of the dispersed phase increases, the flow resistance at the neck of the droplet increases, reducing the flow towards the droplet, hence contributing to produce smaller droplet size too~\cite{Lan15}.

\subsection{Influence of geometrical parameters}
\label{Geometrical parameters}
\begin{figure*}[ht]
	\centering 
        \includegraphics[width=1.08\textwidth]{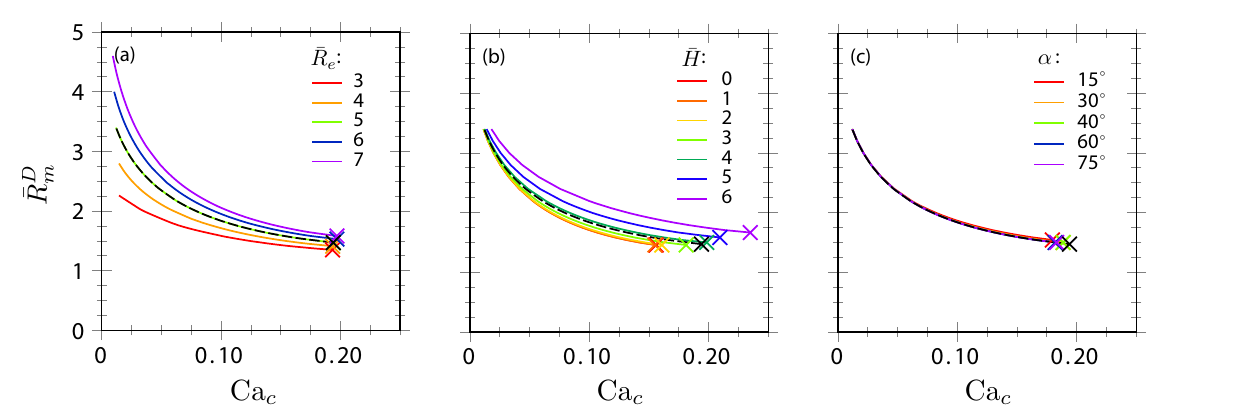}
	\caption{Influence of the geometrical parameters on the droplet size for $\lambda=1/23$. Parameters (a) $\bR_{e}$, (b) $\bar{H}$ and (c) $\alpha$ are varied around the nominal values of couple 1 represented in dashed black line.}  
	\label{GeoInf}
\end{figure*}

The quasi-static approach is now used to study the influence of the geometrical parameters $\bR_{e}$, $\bar{H}$ and $\alpha$ (see \figref{Sketch}) as compared to the nominal values corresponding to the geometry of couple 1 (see Table~\ref{tbl:data}), with $\lambda=1/23$. It can be observed in \figref{GeoInf}(a) that a smaller extraction capillary leads to smaller drops for the same $\Ca_c$, since the viscous stresses exerted by the continuous phase on the meniscus obviously increases with the confinement. On the contrary, the dripping-jetting transition at $\Ca_c^*$ (crosses) is almost insensitive to the radius of the extraction capillary. 
\figref{GeoInf}(b) shows that the meniscus equivalent radius is quite insensitive to $\bar{H}$ for $\bar{H}<2$, above which a larger value of $\bar{H}$ leads to a larger meniscus. It can be explained by the flow field coming from an infinite medium inside a tube in absence of the dispersed phase and the nozzle, where the velocity is uniform inside the extraction capillary but decays in the axis of symmetry as moving away from the entrance. The length scale of this effect is the radius of the capillary. For larger $\bar{H}$, the flow rate should be higher to keep the same stresses around the meniscus, and thus the same droplet size. Interestingly, the dripping-jetting transition (crosses) occurs for large $\Ca_c^*$ as $\bar H$ is increased. Next, no influence of $\alpha$ is found for the meniscus equivalent radius but small and non-monotonic influence on $\Ca_c^*$ is revealed, with a maximum around $50^\circ$, as shown in \figref{GeoInf}(c). Note finally that no significant influence of the thickness of the extraction tube wall, $\bar e$, could be identified, such as this parameter has been disregarded from our parametric analysis.

\subsection{Influence of inertia}
\label{Inertia}
\begin{figure}[ht]
	\centering 
        \includegraphics[width=0.4\textwidth]{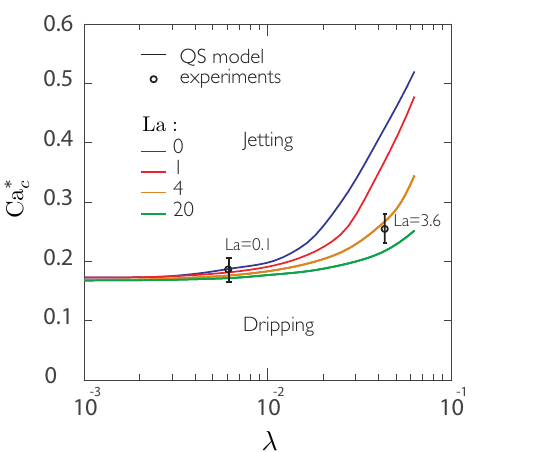}
	\caption{Influence of $\lambda$ on the quasi-static dripping-jetting transition at $\Ca_c^*$ for different values of the Laplace number $\La$ gauging the role of inertia. The geometry of couple 2 has been used.}  
	\label{fig:inertia}
\end{figure}

As mentioned in section~\ref{Modelling}, all results for the drop size have been obtained with the Stokes equation, i.e. with $\La=0$, since no influence of inertia were found for $\La$ up to 100 on the drop size predictions. Nevertheless, we found that inertia can still significantly influence the occurrence of the quasi-static dripping-jetting transition. \figref{fig:inertia} shows that inertia effects enhance the dripping-jetting transition, i.e. $\Ca_c^*$ decreases with increasing $\La$ for a fixed $\lambda$, and that the enhancement is more pronounced as $\lambda$ increases. 
On the contrary, inertia plays almost no role in the inviscid limit ($\lambda \rightarrow 0$). 

Experimentally, this transition could only be identified with the geometry of couple 2 and the fluid properties given in Table~\ref{tbl:data}. The corresponding point is shown in \figref{fig:inertia} for $\lambda=1/23\approx 4\times 10^{-2}$ and $\La=3.6$, with the error bar covering the range in $\Ca_c$ over which the transition was identified (see \figref{Fig:DripJet}). Despite the large error bar, the agreement with the quasi-static approach including inertia (see orange line for $\La=4$ in \figref{fig:inertia}) is satisfactory and confirms the important role of inertia for viscous droplets. Indeed, for the case of less viscous droplets, another pair of fluids corresponding to $\lambda\approx 6\times 10^{-3}$ and $\La=0.1$ has also been tested experimentally, showing again an excellent agreement with the modelling, and demonstrating a much lower influence of inertia effects for low viscosity ratios.

\section{Discussions and conclusions}
\label{Conclusions}
In this paper, we have presented the Raydrop device allowing for the generation of droplets in an axisymmetric flow-focusing, that we have renamed non-embedded {\it co-flow-focusing}. This configuration, together with the quasi-static approach, has enabled a systematic parametric analysis, considering the influence of all relevant parameters. In the quasi-static dripping regime, droplet size has been shown (i) to decrease with increasing the flow rate of the continuous phase, (ii) to decrease with increasing the viscosity ratio between the dispersed phase and the continuous phase, (iii) to increase with increasing the ratio of the radii between the extraction capillary and the nozzle, (iv) to increase with the increasing inter-distance between the nozzle and the extraction capillary, all these dependencies being related to the viscous forces associated to the droplet detachment, like the gravity force in pendant droplets. On the contrary, the droplet size has been found to be rather insensitive to inertia and to the inclination angle of the nozzle, even though these parameters have been shown to significantly affect the quasi-static dripping to jetting transition.

Besides the excellent agreement obtained between experiments and simulations, not only in the quasi-static limit but also in transient, we have experimentally identified the dripping to jetting transition when increasing the dispersed flow rate ($Q_d$) for a given continuous flow rate ($Q_c$). This transition has however not been explored in detail using our simulations as it requires full transient parametric analysis, hence much more computing resources than for simulating steady flows. Another justification for having disregarded a full parametric analysis in transient for the dripping regime is that the experimental droplet size has been shown to vary only slightly with $Q_d$ at a fixed $Q_c$. This characteristic is remarkable as it demonstrates the dominant role of the viscous forces induced by the continuous phase on the droplet formation and it therefore supports the pertinence of the quasi-static approach beyond its validity domain. Consequently, since increasing $Q_d$ has little effect on the droplet size, it essentially modulates the droplet frequency (see \figref{Fig:WO30mu}).

\begin{figure*}[ht!] 
        \centering 
        \includegraphics[width=\textwidth]{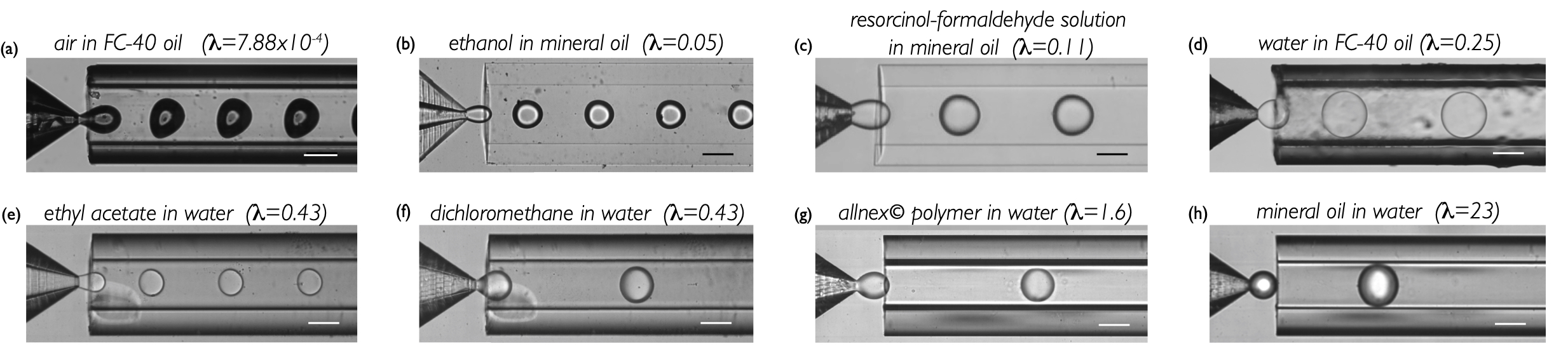}
        \caption{Experimental images of droplet generation in the dripping regime using the Raydrop with different couples and various pairs of fluids in a wide range of viscosity ratios. The scale bar is 100 $\mu$m. }
        \label{Fig:Pannel}
\end{figure*}

While this study has been performed for water in oil, we illustrate in Fig.~\ref{Fig:Pannel} the generation of bubbles and droplets involving a wide variety of pair of fluids. 
This figure demonstrates the universality of the Raydrop to operate properly independently of the wetting properties of the materials in contact with the fluids, and independently of the physico-chemical properties of these fluids (interfacial tension, viscosity, density, miscibility). 
It additionally indicates that tuning the diameters of the nozzle tip and/or the extraction capillary enables to cover a wide range of droplet diameters, theoretically from 20 to 400$\,\mu$m, with any given couple of fluids, a feature which is not achievable today with any other single device available.

In the context of a growing demand of controlled droplets in many areas, the Raydrop emerges therefore as a very robust and versatile solution easily implementable in laboratories with little experience and facilities in microfluidics.

\section*{Conflicts of interest}
The results presented in this paper are covered by a patent application of the Universit\'e libre de Bruxelles~\cite{Scheid19}.

\section*{Acknowledgements}
This work was financially supported by the First-Spin-Off program of the Walloon region as well as by the Fonds de la Recherche Scientifique-F.N.R.S. (Postdoctoral Researcher position of B. Sobac and Senior Research Associate position of B. Scheid). We thank Lionel Matthys for having suggested the terminologies {\it co-flow-focusing}, as well as {\it Raydrop}, in memory of Raymond Brandt who assisted us in the design.

\footnotesize{
\bibliography{Raydrop_arxiv.bib}

\begin{thebibliography}{10}

\bibitem{Gunther06}
Axel Gunther and Klavs~F. Jensen.
\newblock Multiphase microfluidics: from flow characteristics to chemical and
  materials synthesis.
\newblock {\em Lab Chip}, 6:1487--1503, 2006.

\bibitem{Teh08}
Shia-Yen Teh, Robert Lin, Lung-Hsin Hung, and Abraham~P. Lee.
\newblock Droplet microfluidics.
\newblock {\em Lab Chip}, 8:198--220, 2008.

\bibitem{Tran13}
TM~Tran, F~Lan, CS~Thompson, and AR~Abate.
\newblock From tubes to drops: droplet-based microfluidics for
  ultrahigh-throughput biology.
\newblock {\em Journal of Physics D: Applied Physics}, 46(11):114004, 2013.

\bibitem{Anna16}
Shelley~Lynn Anna.
\newblock Droplets and {Bubbles} in {Microfluidic} {Devices}.
\newblock {\em Annual Review of Fluid Mechanics}, 48(1):285--309, 2016.

\bibitem{Shang17}
Luoran Shang, Yao Cheng, and Yuanjin Zhao.
\newblock Emerging droplet microfluidics.
\newblock {\em Chemical Reviews}, 117(12):7964--8040, 2017.

\bibitem{Bremond12}
Nicolas Bremond and J\'er\^ome Bibette.
\newblock Exploring emulsion science with microfluidics.
\newblock {\em Soft Matter}, 8:10549--10559, 2012.

\bibitem{Candoni19}
Nadine Candoni, Romain Grossier, Mehdi Lagaize, and St{\'e}phane Veesler.
\newblock Advances in the use of microfluidics to study crystallization
  fundamentals.
\newblock {\em Annual Review of Chemical and Biomolecular Engineering},
  10(1):59--83, 2019.

\bibitem{Song06}
Helen Song, Delai~L. Chen, and Rustem~F. Ismagilov.
\newblock Reactions in droplets in microfluidic channels.
\newblock {\em Angewandte Chemie International Edition}, 45(44):7336--7356,
  2006.

\bibitem{Lee17}
Doojin Lee, Cifeng Fang, Aniket~S. Ravan, Gerald~G. Fuller, and Amy~Q. Shen.
\newblock Temperature controlled tensiometry using droplet microfluidics.
\newblock {\em Lab Chip}, 17:717--726, 2017.

\bibitem{Conchouso14}
D.~Conchouso, D.~Castro, S.~A. Khan, and I.~G. Foulds.
\newblock Three-dimensional parallelization of microfluidic droplet generators
  for a litre per hour volume production of single emulsions.
\newblock {\em Lab Chip}, 14:3011--3020, 2014.

\bibitem{Nunes13}
J~K Nunes, S~S~H Tsai, J~Wan, and H~A Stone.
\newblock Dripping and jetting in microfluidic multiphase flows applied to
  particle and fibre synthesis.
\newblock {\em Journal of Physics D: Applied Physics}, 46(11):114002, feb 2013.

\bibitem{Baccouche17}
Alexandre Baccouche, Shu Okumura, Rémi Sieskind, Elia Henry, Nathanaël
  Aubert-Kato, Nicolas Bredeche, Jean-François Bartolo, Valérie Taly, Yannick
  Rondelez, Teruo Fujii, and Anthony~J. Genot.
\newblock Massively parallel and multiparameter titration of biochemical assays
  with droplet microfluidics.
\newblock {\em Nature Protocols}, 12:1912--1932, 2017.

\bibitem{Ali-Cherif12}
Ana\"is Ali-Cherif, Stefano Begolo, St\'ephanie Descroix, Jean-Louis Viovy, and
  Laurent Malaquin.
\newblock Programmable magnetic tweezers and droplet microfluidic device for
  high-throughput nanoliter multi-step assays.
\newblock {\em Angewandte Chemie International Edition}, 51(43):10765--10769,
  2012.

\bibitem{Zhang16}
Yonghao Zhang and Hui-Rong Jiang.
\newblock A review on continuous-flow microfluidic pcr in droplets: Advances,
  challenges and future.
\newblock {\em Analytica Chimica Acta}, 914:7 -- 16, 2016.

\bibitem{Agresti10}
Jeremy~J. Agresti, Eugene Antipov, Adam~R. Abate, Keunho Ahn, Amy~C. Rowat,
  Jean-Christophe Baret, Manuel Marquez, Alexander~M. Klibanov, Andrew~D.
  Griffiths, and David~A. Weitz.
\newblock Ultrahigh-throughput screening in drop-based microfluidics for
  directed evolution.
\newblock {\em Proceedings of the National Academy of Sciences},
  107(9):4004--4009, 2010.

\bibitem{Damiati18}
S.~Damiati, U.~B. Kompella, S.~A. Damiati, and R.~Kodzius.
\newblock Microfluidic devices for drug delivery systems and drug screening.
\newblock {\em Genes}, 9(2):103, 2016.

\bibitem{Klein15}
A.M. Klein, L.~Mazutis, I.~Akartuna, N.~Tallapragada, A.~Veres, V.~Li,
  L.~Peshkin, D.A. Weitz, and M.W. Kirschner.
\newblock Droplet barcoding for single cell transcriptomics applied to
  embryonic stem cells.
\newblock {\em Cell}, 161(5):1187--1201, 2015.

\bibitem{Brouzes09}
Eric Brouzes, Martina Medkova, Neal Savenelli, Dave Marran, Mariusz Twardowski,
  J.~Brian Hutchison, Jonathan~M. Rothberg, Darren~R. Link, Norbert Perrimon,
  and Michael~L. Samuels.
\newblock Droplet microfluidic technology for single-cell high-throughput
  screening.
\newblock {\em Proceedings of the National Academy of Sciences},
  106(34):14195--14200, 2009.

\bibitem{Ryckelynck15}
M.~Ryckelynck, S.~Baudrey, C.~Rick, A.~Marin, F.~Coldren, E.~Westhof, and A.~D.
  Griffiths.
\newblock Using droplet-based microfluidics to improve the catalytic properties
  of rna under multiple-turnover conditions.
\newblock {\em RNA}, 21:458--469, 2015.

\bibitem{Guillot07}
Pierre Guillot, Annie Colin, Andrew~S. Utada, and Armand Ajdari.
\newblock Stability of a {Jet} in {Confined} {Pressure}-{Driven} {Biphasic}
  {Flows} at {Low} {Reynolds} {Numbers}.
\newblock {\em Physical Review Letters}, 99(10):104502, 2007.

\bibitem{Serra07}
Christophe Serra, Nicolas Berton, Michel Bouquey, Laurent Prat, and Georges
  Hadziioannou.
\newblock A {Predictive} {Approach} of the {Influence} of the {Operating}
  {Parameters} on the {Size} of {Polymer} {Particles} {Synthesized} in a
  {Simplified} {Microfluidic} {System}.
\newblock {\em Langmuir}, 23(14):7745--7750, 2007.

\bibitem{Utada05}
A.~S. Utada, E.~Lorenceau, D.~R. Link, P.~D. Kaplan, H.~A. Stone, and D.~A.
  Weitz.
\newblock Monodisperse {Double} {Emulsions} {Generated} from a {Microcapillary}
  {Device}.
\newblock {\em Science}, 308(5721):537--541, 2005.

\bibitem{Benson13}
Bryan~R. Benson, Howard~A. Stone, and Robert~K. Prud'homme.
\newblock An €œoff-the-shelf€ capillary microfluidic device that
  enables tuning of the droplet breakup regime at constant flow rates.
\newblock {\em Lab on a Chip}, 13(23):4507--4511, 2013.

\bibitem{Bandulasena19}
Monalie~V. Bandulasena, Goran~T. Vladisavljevi{\'c}‡, and Brahim Benyahia.
\newblock Versatile reconfigurable glass capillary microfluidic devices with
  lego\textregistered inspired blocks for drop generation and micromixing.
\newblock {\em Journal of Colloid and Interface Science}, 542:23 -- 32, 2019.

\bibitem{Evangelio16}
A.~Evangelio, F.~Campo-Cortès, and J.~M. Gordillo.
\newblock Simple and double microemulsions via the capillary breakup of highly
  stretched liquid jets.
\newblock {\em Journal of Fluid Mechanics}, 804:550--577, 2016.

\bibitem{Cruz-Mazo16}
Francisco Cruz-Mazo, JM~Montanero, and AM~Ga{\~n}{\'a}n-Calvo.
\newblock Monosized dripping mode of axisymmetric flow focusing.
\newblock {\em Physical Review E}, 94(5):053122, 2016.

\bibitem{Guillot08}
Pierre Guillot, Annie Colin, and Armand Ajdari.
\newblock Stability of a jet in confined pressure-driven biphasic flows at low
  reynolds number in various geometries.
\newblock {\em Physical Review E}, 78(1):016307, 2008.

\bibitem{Cordero11}
Mar{\'\i}a~Luisa Cordero, Fran{\c{c}}ois Gallaire, and Charles~N Baroud.
\newblock Quantitative analysis of the dripping and jetting regimes in
  co-flowing capillary jets.
\newblock {\em Physics of Fluids}, 23(9):094111, 2011.

\bibitem{Richards95}
John~R Richards, Antony~N Beris, and Abraham~M Lenhoff.
\newblock Drop formation in liquid--liquid systems before and after jetting.
\newblock {\em Physics of Fluids}, 7(11):2617--2630, 1995.

\bibitem{Hua07}
Jinsong Hua, Baili Zhang, and Jing Lou.
\newblock Numerical simulation of microdroplet formation in coflowing
  immiscible liquids.
\newblock {\em AIChE Journal}, 53(10):2534--2548, 2007.

\bibitem{Lan15}
Wenjie Lan, Shaowei Li, and Guangsheng Luo.
\newblock Numerical and experimental investigation of dripping and jetting flow
  in a coaxial micro-channel.
\newblock {\em Chemical Engineering Science}, 134:76--85, 2015.

\bibitem{Bai17}
Feng Bai, Xiaoming He, Xiaofeng Yang, Ran Zhou, and Cheng Wang.
\newblock Three dimensional phase-field investigation of droplet formation in
  microfluidic flow focusing devices with experimental validation.
\newblock {\em International Journal of Multiphase Flow}, 93:130--141, 2017.

\bibitem{Wu08}
Long Wu, Michihisa Tsutahara, Lae~Sung Kim, and ManYeong Ha.
\newblock Three-dimensional lattice boltzmann simulations of droplet formation
  in a cross-junction microchannel.
\newblock {\em International Journal of Multiphase Flow}, 34(9):852--864, 2008.

\bibitem{Zhou06}
Chunfeng Zhou, Pengtao Yue, and James~J Feng.
\newblock Formation of simple and compound drops in microfluidic devices.
\newblock {\em Physics of Fluids}, 18(9):092105, 2006.

\bibitem{Garstecki05}
Piotr Garstecki, Howard~A Stone, and George~M Whitesides.
\newblock Mechanism for flow-rate controlled breakup in confined geometries: A
  route to monodisperse emulsions.
\newblock {\em Physical Review Letters}, 94(16):164501, 2005.

\bibitem{Li15}
Z~Li, AM~Leshansky, LM~Pismen, and P~Tabeling.
\newblock Step-emulsification in a microfluidic device.
\newblock {\em Lab on a Chip}, 15(4):1023--1031, 2015.

\bibitem{Boucher75}
E.~A. Boucher, M.~J.~B. Evans, and Frederick~Charles Frank.
\newblock Pendent drop profiles and related capillary phenomena.
\newblock {\em Proceedings of the Royal Society of London. A. Mathematical and
  Physical Sciences}, 346(1646):349--374, 1975.

\bibitem{Valet18}
M~Valet, L-L Pontani, AM~Prevost, and E~Wandersman.
\newblock Quasistatic microdroplet production in a capillary trap.
\newblock {\em Physical Review Applied}, 9(1):014002, 2018.

\bibitem{Basaran90}
Osman~A Basaran and LE~Scriven.
\newblock Axisymmetric shapes and stability of pendant and sessile drops in an
  electric field.
\newblock {\em Journal of Colloid and Interface Science}, 140(1):10--30, 1990.

\bibitem{Cristini04}
Vittorio Cristini and Yung-Chieh Tan.
\newblock Theory and numerical simulation of droplet dynamics in complex flows
  - a review.
\newblock {\em Lab on a Chip}, 4(4):257--264, 2004.

\bibitem{Martinez20}
A.~Martínez-Calvo, J.~Rivero-Rodriguez, B.~Scheid, and A.~Sevilla.
\newblock Natural break-up and satellite formation regimes of surfactant-laden
  liquid threads.
\newblock {\em Journal of Fluid Mechanics}, 883:A35, 2020.

\bibitem{VanBrummelen01}
EH~Van~Brummelen, HC~Raven, and B~Koren.
\newblock Efficient numerical solution of steady free-surface navier--stokes
  flow.
\newblock {\em Journal of Computational Physics}, 174(1):120--137, 2001.

\bibitem{Rivero20}
Javier Rivero-Rodriguez, Miguel Perez-Saborid, and Benoit Scheid.
\newblock Pdes on deformable domains: Boundary arbitrary lagrangian-eulerian
  (bale) and deformable boundary perturbation (dbp) methods.
\newblock {\em arXiv preprint arXiv:1810.10001}, 2018.

\bibitem{Rivero18}
Javier Rivero-Rodriguez and Benoit Scheid.
\newblock Bubble dynamics in microchannels: inertial and capillary migration
  forces.
\newblock {\em Journal of Fluid Mechanics}, 842:215--247, 2018.

\bibitem{Rivero19}
Javier Rivero-Rodriguez and Benoit Scheid.
\newblock Mass transfer around bubbles flowing in cylindrical microchannels.
\newblock {\em Journal of Fluid Mechanics}, 869:110--142, 2019.

\bibitem{Erb11}
Randall~M. Erb, Dominik Obrist, Philipp~W. Chen, Julia Studer, and Andre~R.
  Studart.
\newblock Predicting sizes of droplets made by microfluidic flow-induced
  dripping.
\newblock {\em Soft Matter}, 7(19):8757, 2011.

\bibitem{Suryo06}
Ronald Suryo and Osman~A Basaran.
\newblock Tip streaming from a liquid drop forming from a tube in a co-flowing
  outer fluid.
\newblock {\em Physics of Fluids}, 18(8):082102, 2006.

\bibitem{Scheid19}
B.~Scheid, A.~Dewandre, and Y.~Vitry.
\newblock Droplet and/or bubble generator, ~10 2019.
\newblock PCT/EP2018/067960.

\end{thebibliography}
\bibliographystyle{unsrt}
}

\end{document}